\documentclass[11pt]{extarticle}
\usepackage{makecell}
\usepackage[toc,page]{appendix}
\usepackage[toc,page]{appendix}
\usepackage[affil-it]{authblk}
\usepackage{lineno,hyperref}
\usepackage{forest}
\modulolinenumbers[50]
\usepackage{amsmath}
\usepackage{graphicx}
\usepackage{dcolumn}
\usepackage{bm}
\usepackage{color}
\usepackage{caption}
\usepackage{ragged2e}
\usepackage{tabularx,ragged2e,booktabs,caption}
\usepackage{amsmath}
\usepackage{algorithm}
\usepackage{algpseudocode}
\usepackage{bbm}
\usepackage{dsfont}
\usepackage{mathrsfs}
\usepackage{forest,adjustbox}
\usepackage{tikz}
\makeatletter
\def\BState{\State\hskip-\ALG@thistlm}
\makeatother
\usepackage{ctable}
\usepackage{xcolor}
\usepackage{algorithm}
\usepackage{geometry}
\usepackage{makecell}
\usepackage{boldline}
\usepackage[utf8]{inputenc}
\usepackage[italian, english]{babel}

\usepackage{helvet}
\usepackage{mathtools} 
\usepackage{amsmath}

\usepackage{amssymb}
\usepackage{amsthm}
\usepackage{mathrsfs}
\usepackage{amsfonts}
\usepackage{numprint}
\usepackage{verbatim}
\usepackage{algorithm}
\usepackage{graphicx}

\usepackage{hyperref}
\usepackage{bm}
\usepackage{booktabs}
\usepackage{eucal}
\usepackage{rotating}
\usepackage[pdftex]{lscape}
\usepackage{longtable}
\usepackage[italian]{varioref}
\usepackage{subfigure}
\usepackage{color}
\usepackage{algorithm}
\usepackage{algpseudocode}
\usepackage{bbm}
\usepackage{dsfont}
\makeatletter
\def\BState{\State\hskip-\ALG@thistlm}
\makeatother
\usepackage{tikz}
\usetikzlibrary{3d}
\usetikzlibrary{patterns}
\usetikzlibrary{calc}
\usetikzlibrary{arrows}
\usepackage{fancyhdr}

\usepackage{pgfplots}
\usepackage{pgfplotstable}
\usepackage{etoolbox}
\usetikzlibrary{decorations.pathreplacing}
\makeatletter
\newcommand{\gettikzxy}[3]{%
	\tikz@scan@one@point\pgfutil@firstofone#1\relax
	\edef#2{\the\pgf@x}%
	\edef#3{\the\pgf@y}%
}
\makeatother

\usepackage{multirow}

\usepackage{tabularx,ragged2e,booktabs,caption}
\usepackage{etoolbox}
\usepackage{adjustbox}
\usepackage{keyval}

\makeatletter
\define@key{cascading}{width}{\cascading@wd=#1}
\define@key{cascading}{sep}{\def\cascading@sep{#1}}
\newdimen\cascading@wd

\newcommand{\cascadingblocks}[2][]{%
	\setkeys{cascading}{sep=2ex,#1}%
	\leavevmode\vbox{\offinterlineskip
		\@for\next:=#2\do{%
			\settowidth{\dimen@}{\next}%
			\ifdim\dimen@>\cascading@wd
			\cascading@wd=\dimen@
			\fi
		}%
		\@for\next:=#2\do{%
			\cascading@rule
			\hbox{\fbox{\hb@xt@\cascading@wd{\hss\next\hss}}}%
		}
	}
}
\def\cascading@rule{%
	\def\cascading@rule{%
		\hb@xt@\dimexpr\cascading@wd+2\fboxsep+2\fboxrule\relax
		{\hss\vrule\@height\cascading@sep\hss}%
	}%
}
\makeatother

\newcommand{\M}[2][]{{\bm{#1\mathbf{\MakeUppercase{#2}}}}} 
\newcommand{\T}[2][]{\boldsymbol{#1\mathscr{\MakeUppercase{#2}}}} 
\newcommand{\Mz}[3][]{\M[#1]{#2}_{(#3)}}

\newcommand{\argmin}{\arg\!\min}

\newcommand{\R}{\mathbb{R}}

\DeclareMathOperator{\E}{\mathbb{E}}

\usepackage{fancyhdr}
\begin{document}
	
	\title{\Large Unveil stock correlation via a new tensor-based decomposition method}
	\vspace{-5pt}
	\author{Giuseppe Brandi\thanks{Corresponding author. Email: giuseppe.brandi@kcl.ac.uk} $^{,\dagger}$, Ruggero Gramatica$^{\ddagger}$, T. Di Matteo$^{\dagger,\S,\P}$}
	
	\affil{$\dagger$ Department of Mathematics, King's College London, The Strand, London, WC2R 2LS, UK\\
		$\ddagger$ Yewno, 499 Hamilton Ave, Palo Alto, California 94301, USA\\
		$\S$ Department of Computer Science, University College London, Gower Street, London, WC1E 6BT, UK\\
		$\P$ Complexity Science Hub Vienna, Josefstaedter Strasse 39, A 1080 Vienna, Austria}
	
	\providecommand{\keywords}[1]{\textbf{\textit{Keywords:}} #1}

	\date{}
	
	\maketitle
	
	\vspace{-15pt}
	
	\begin{abstract}
		Portfolio allocation and risk management make use of correlation matrices and heavily rely on the choice of a proper correlation matrix to be used. In this regard, one important question is related to the choice of the proper sample period to be used to estimate a stable correlation matrix.
		This paper addresses this question and proposes a new methodology to estimate the correlation matrix which doesn't depend on the chosen sample period. 
		This new methodology is based on tensor factorization techniques. In particular, combining and normalizing factor components, we build a correlation matrix which shows emerging structural dependency properties not affected by the sample period. To retrieve the factor components, we propose a new tensor decomposition (which we name Slice-Diagonal Tensor (SDT) factorization) and compare it to the two most used tensor decompositions, the Tucker and the PARAFAC. We have that the new factorization is more parsimonious than the Tucker decomposition and more flexible than the PARAFAC.
		Moreover, this methodology applied to both simulated and empirical data shows results which are robust to two non-parametric tests, namely Kruskal-Wallis and Kolmogorov-Smirnov tests. Since the resulting correlation matrix features stability and emerging structural dependency properties, it can be used as alternative to other correlation matrices type of measures, including the Person correlation.

	\end{abstract}
	

	\section{Introduction}
	Since the pioneering work by Markowitz \cite{Markowitz59}, correlation matrices are ubiquitous in finance,  both in academia and industry. Pearson correlation is the most used measure of association between stock returns \cite{aste2010correlation,pozzi2012exponential} because of its simple application and interpretation. However, many other types of correlation measures can be studied \cite{musmeci2017multiplex}. In fact, portfolio construction and risk management heavily rely on correlation matrices  \cite{mantegna1999, bartolozzi1, tumminello2010}. In particular, when the structural (not time dependent) features of the correlation matrix are necessary, i.e. financial contagion, cross-assets VaR forecasting and long term investments for which it is not possible to frequently rebalance the portfolio, estimating a representative correlation matrix becomes essential. Nevertheless, this has been proven to be difficult to achieve. The main problem which arises in this context is the selection of the appropriate sample period. Heuristic approaches to tackle the problem exist, e.g. taking very long samples, but all of them still rely on the sample period selected. The potential issue of these methodologies is that dependences can be netted out (or inflated) if samples with different market conditions are employed. For example, companies with similar supply chains are usually correlated (and they should be), but if we take a sample in which for a period of time one specific market factor (which affect only one of the two companies) has a huge downturn, the whole correlation can be destroyed even if it is caused by an exogenous, transient market factor. On the contrary, companies operating in very different markets, should be moderately or negatively correlated. However, if a financial crisis is in the sample,\footnote{During financial crisis, stocks tend to correlate since stock prices falls altogether.} the correlation would be inflated even if their structural dependency is low or even negative. 
	In the literature we can find different approaches in analysing correlation matrices and their stability. Some papers analyse the sample size necessary for the correlation matrix to stabilize \cite{Schonbrodt2013} while others test if in different periods the correlation matrix remains statistically similar \cite{gordon1998, kao2018}. Moreover, it is clear from empirical evidence that correlation matrices change over time, and respond to market fluctuations \cite{engel1999}. In particular, forecasting Value at Risk (VaR) for  risk management across assets needs the correlation and covariance matrix to be as stable as possible, in order to be able to represent future risk and possible risk spillovers \cite{engel1999}. Regarding portfolio construction and its analysis, it is well known that ex-ante optimal mean-variance efficient portfolio construction requires correlation and covariance structures to be stable over time \cite{lee1998,piet1996,cheung1991}. In fact, only if the dependency structure is inter-temporally invariant ex-post portfolio analysis gives insightful results. The necessity of having stable correlations for risk management and portfolio analysis is clear. In order to answer to this important question, in this paper we introduce a methodology based on tensor factorization \cite{bro1998multi, henrion1994n, kroonenberg2008applied, grasedyck2013literature, kolda2009tensor, anandkumar2014tensor, faber2003recent}. In particular, tensor decomposition is employed to factorize the $3$-rd order covariance tensor, formed by the time series of covariance matrices. To achieve this, there are different decompositions we can use to factorize a tensor. In this paper we propose a new factorization technique (which we name Slice-Diagonal Tensor (SDT) decomposition) and compare it to the two most important factorization methods, i.e. PARAFAC \cite{harshman1970foundations,carroll1970analysis} and Tucker decomposition \cite{tucker1963implications,tucker1964extension,tucker1966some}. This new decomposition method represents, similarly to the Tucker decomposition, a more flexible version of the PARAFAC model, while being more parsimonious than it. This allows to handle huge-dimensional problems with fewer elements, which helps reducing over-fitting and highly correlated solutions. The tensor decomposition permits to factorize the covariance tensor into static and dynamic components. The static components represent the hidden, long-run, behaviour of the interdependences between stocks, while the dynamic component represents the time dependent structure in the latent space, which is a proxy of market variance dynamic. Combining the static components, we build a matrix that after being suitably normalized, represents a correlation matrix, which we name hidden correlation matrix (HCM). This methodology projects the interdependencies into a lower-dimensional space and hence it gives rise to a low-rank approximation of the true hidden correlation matrix. In addition, if one needs to forecast the correlation matrix at one specific time period in the future, it would need only to forecast the dynamic component instead of the whole correlation matrix. This makes the decomposition also a parsimonious way to forecast the full correlation matrix. 
	To the best of our knowledge, this is the first time a structural correlation matrix is built instead of only tested for stability. To provide evidence of the structural dependencies of the hidden correlation matrix, we show, using both visual inspection and two statistical tests based on eigenvalue distributions, that the hidden correlation matrix dependency structure is statistically time invariant. This result shows that structural dependencies between stocks are retrieved by the top-down hidden correlation construction. Since this correlation matrix relies only weakly on the selected sample period, it is a good proxy for forecasting future correlations. This result unveils that hidden correlation matrix can be used as an alternative to the standard correlation matrix, in particular when long-run stability is required. The paper is structured as follows. Section \ref{decomp_sec} is devoted to the full methodology introduced in this paper to unveil the structural correlation matrix. We present all the steps necessary to go from the input covariance tensor to the output correlation matrix using the proposed SDT decomposition and alternative tensor factorizations. Sections \ref{sim_sec} and \ref{app_sec} are dedicated to a simulation and an empirical application of the methods introduced in section \ref{decomp_sec}  where we show the performances of the SDT decomposition compared to PARAFAC and Tucker and how the algorithmic procedure is able to efficiently retrieve the structural correlation matrix when applied to data. Finally, section \ref{conclusion_sec} concludes.

	\section{Hidden correlation construction}\label{decomp_sec}
	In this section, we explain the full methodology used to build the hidden correlation matrix (HCM). We start by introducing the tensor notation and the various factorizations, focusing on the proposed SDT decomposition. We then discuss model complexity and information criteria \cite{schwarz1978,akaike1974,CAVANAUGH1997} employed to find the optimal number of components for each factorization technique. Finally, we show how to combine and normalize the factor components to produce the hidden correlation matrix, which is the final output of our algorithmic methodology.

\subsection{Tensor decomposition}	
\begin{figure}[H]		
		\centering	
		\includegraphics[width=0.3\linewidth,height=0.2\textheight]{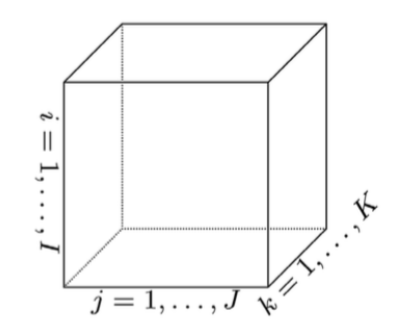}		
		\caption{Illustration of a $3$-rd order tensor of dimensions $I \times J \times K$.}	
		\label{fig:tt1}	
	\end{figure}

	Tensors are a generalization of scalars, vectors and matrices and their \textit{order} defines them \cite{kroonenberg2008applied, kolda2009tensor, anandkumar2014tensor}. The \textit{order} of a tensor, which is the number of dimensions that characterises them, is also referred to as ways or modes. Scalars are $0$-th order tensors, vectors are $1$-st order tensors and matrices are $2$-nd order tensors. Arrays with more than two dimensions are referred to as higher order tensors or just as tensors. Throughout this paper, the notation will follow the standard convention established in \cite{kolda2009tensor, bader2012matlab, anandkumar2014tensor}: $x$ is a scalar, $\textbf{x}$ is a vector, $\textbf{X}$ is a matrix and $\T{X}$ is a tensor. Figure \ref{fig:tt1} represents an example of a $3$-rd order tensor with dimensions $I \times J \times K$.

In the case of a covariance, correlation or network tensor, the first two dimensions would be dedicated to the dependency matrix, while the third dimension is related to the time component. In particular, for a covariance tensor composed of $M$ stocks and $T$ time stamps, we would have $I=J=M$ and $K=T$. A tensor, as in the case of factorization for matrices \cite{Jolliffe, matrix_factorization}, can be decomposed in smaller objects \cite{anandkumar2014tensor,bro1998multi,kolda2009tensor}. These objects contain information of a specific dimension of the $3$-rd order tensor which in the case of a time series of matrices, representing correlations or covariances, are related to the the cross-sectional, static, dimensions and to the time dimension. More specifically, the first two factor matrices represent the static (long-run) interconnections between stocks, while the third factor component represents the dynamic factor of the market variance. The former, when combined, form a matrix that when normalized, represents the hidden correlation matrix. To factorize a tensor, there are different decompositions available from the statistical toolbox. We rely on the two most important factorization methods, i.e. PARAFAC \cite{harshman1970foundations,carroll1970analysis} and Tucker decomposition \cite{tucker1963implications,tucker1964extension,tucker1966some}. We briefly recall those decompositions in sections \ref{parafac_sec} and \ref{tucker_sec}. Further, in section \ref{sdt_sec} we propose and explore a new factorization method which we name Slice-Diagonal Tensor (SDT) decomposition. This new factorization strategy is a trade-off between parsimony and feasibility. In particular, it represents a generalization of the PARAFAC model as well as a parsimonious version of the Tucker decomposition. 
	
	\subsection{PARAFAC decomposition}\label{parafac_sec}
\begin{figure}[H]
		\centering
		\begin{tikzpicture}[scale=.7,namenode/.style={scale=.85}]

		\def\ix{3} %
		\def\iy{3} %
		\def\iz{2.5} %

		\def\corescale{1.75}
		\def\rx{\ix/\corescale}
		\def\ry{\iy/\corescale}
		\def\rz{\iz/\corescale}

		\coordinate (XFrontLowerLeft) at (0,0);
		\draw (XFrontLowerLeft) rectangle ++ (\ix,\iy); %
		\begin{scope}[shift={(XFrontLowerLeft)},canvas is zx plane at y=\iy,rotate=90]
		\draw (0,0) rectangle ++ (\ix,\iz); %
		\end{scope}
		\begin{scope}[shift={(XFrontLowerLeft)},canvas is zy plane at x=\ix,rotate=90]
		\draw (0,0) rectangle ++ (\iy,\iz); %
		\end{scope}
		\node[namenode] at ($(XFrontLowerLeft) + (0.5*\ix, 0.5*\iy)$)  {\huge{$\T{X}$}};

		\coordinate (ApproxCtr) at ($(XFrontLowerLeft) + (\ix+0.4*\iz,0.75*\iy) + (0.75,0)$);
		\node[namenode] at (ApproxCtr) {$\approx$};

		\coordinate (U1LowerLeft) at ($(ApproxCtr) - (0,0.8*\iy) + (0.75,0.5)$);
		\draw (U1LowerLeft) rectangle ++ (-1.1+\ry,-0.4+\iy);
		\node[namenode] at ($(U1LowerLeft)+(0.2*\ry, 0.4*\iy)$)  {$a_1$};
		
		\coordinate (U2LowerLeft) at ($(U1LowerLeft) + (\rx+\ry-2.75,2)$);
		\draw (U2LowerLeft) rectangle ++ (-0.4+\ix,-1.1+\rx); %
		\node[namenode] at ($(U2LowerLeft)+(0.5*\ix,0.15*\rx)$)  {$b_1$};
		
		\coordinate (U3LowerLeft) at ($(U1LowerLeft) + (-0.,\ry+1)$);
		\begin{scope}[shift={(U3LowerLeft)},canvas is zx plane at y=0,rotate=90]
		\draw (0,0) rectangle ++ (\rz-0.8,\iz-0.1); %
		\end{scope}
		\node[namenode] at ($(U3LowerLeft)+(0.8,0.5)$) {$c_1$};
		
		\coordinate (Plus) at ($(ApproxCtr) + (3.75,-0.15*\iy) + (0.75,0.5)$);
		\node[namenode] at (Plus) {$+$};
		\coordinate (U1LowerLeft2) at ($(Plus) - (0.1,0.8*\iy) + (0.75,0.5)$);
		\draw (U1LowerLeft2) rectangle ++ (-1.1+\ry,-0.4+\iy);
		\node[namenode] at ($(U1LowerLeft2)+(0.2*\ry, 0.4*\iy)$)  {$a_2$};
		
		\coordinate (U2LowerLeft2) at ($(U1LowerLeft2) + (\rx+\ry-2.75,2)$);
		\draw (U2LowerLeft2) rectangle ++ (-0.4+\ix,-1.1+\rx); %
		\node[namenode] at ($(U2LowerLeft2)+(0.5*\ix,0.15*\rx)$)  {$b_2$};
		
		\coordinate (U3LowerLeft2) at ($(U1LowerLeft2) + (-0.,\ry+1)$);
		\begin{scope}[shift={(U3LowerLeft2)},canvas is zx plane at y=0,rotate=90]
		\draw (0,0) rectangle ++ (\rz-0.8,\iz-0.1); %
		\end{scope}
		\node[namenode] at ($(U3LowerLeft2)+(0.8,0.5)$) {$c_2$};
		
		\coordinate (Plus2) at ($(ApproxCtr) + (8.5,-0.15*\iy) + (0.75,0.5)$);
		\node[namenode] at (Plus2) {$+ \dots +$};

		\coordinate (U1LowerLeft3) at ($(Plus2) - (-0.1,0.8*\iy) + (0.75,0.5)$);
		\draw (U1LowerLeft3) rectangle ++ (-1.1+\ry,-0.4+\iy);
		\node[namenode] at ($(U1LowerLeft3)+(0.2*\ry, 0.4*\iy)$)  {$a_R$};
		
		\coordinate (U2LowerLeft3) at ($(U1LowerLeft3) + (\rx+\ry-2.75,2)$);
		\draw (U2LowerLeft3) rectangle ++ (-0.4+\ix,-1.1+\rx); %
		\node[namenode] at ($(U2LowerLeft3)+(0.5*\ix,0.15*\rx)$)  {$b_R$};
		
		\coordinate (U3LowerLeft3) at ($(U1LowerLeft3) + (-0.,\ry+1)$);
		\begin{scope}[shift={(U3LowerLeft3)},canvas is zx plane at y=0,rotate=90]
		\draw (0,0) rectangle ++ (\rz-0.8,\iz-0.1); %
		\end{scope}
		\node[namenode] at ($(U3LowerLeft3)+(0.8,0.5)$) {$c_R$};
		
		\end{tikzpicture}
		\caption{PARAFAC decomposition of a $3$-rd order tensor.}
		\label{fig:parafac}
	\end{figure}
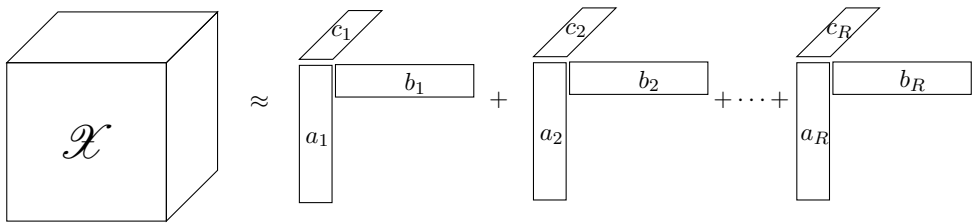

	PARAFAC decomposition was firstly introduced in psychometry by \cite{harshman1970foundations, carroll1970analysis} to study patients for which the variables of interest were observed at different occasions, generating a $3$-rd order tensor in which the dimensions are respectively \textit{I=individuals}, \textit{J=variables} and \textit{K=occasions}. In this case, every resulting factor has information regarding the specific dimension and can be used to cluster patients through time or variables. Another field of research in which the PARAFAC decomposition was successfully used is chemometrics, where it has been used to model fluorescence excitation-emission data. The model was introduced in  chemometrics by \cite{appelof} and then further studied by \cite{bro1997parafac,bro1998multi,anderson1999general,andersen2003practical,bro2006}. Other applications include data mining \cite{acar2005modeling,acar2006}, text analysis and semantic networks \cite{bader2007,bader2008}.The PARAFAC decomposition generalizes the bilinear factor model to the multilinear case. Mathematically, it is based on the idea of rewriting a tensor as the sum of rank-one tensors. Take for example a $3$-rd order tensor $\T{X} \in \R^{I \times J \times K }$. We rewrite this tensor in terms of outer product as:
	\begin{equation}\label{p}
	\T{X} \approx \sum_{r=1}^{R}\textbf{a}_r \circ \textbf{b}_r \circ \textbf{c}_r, 
	\end{equation}
	where $R$ is a positive integer representing the number of components used in the decomposition and $\textbf{a}_r\in \R^{I }$ , $\textbf{b}_r\in \R^{J }$ and $\textbf{c}_r\in \R^{ K}$ are the vector components for $r=1,\dots,R$.   
	 A graphical representation of the PARAFAC decomposition of a $3$-rd order tensor is illustrated in figure \ref{fig:parafac}.

PARAFAC decomposition has the advantage to be generally parsimonious and unique. However, this decomposition is not flexible and can fail to converge to a solution \cite{kolda2009tensor}.

	\subsection{Tucker decomposition}\label{tucker_sec}
	The Tucker decomposition was theorized by Tucker in 1963 \cite{tucker1963implications,tucker1964extension,tucker1966some}. As the PARAFAC decomposition, it represents an extension of the bilinear factor analysis to the multilinear case. It is also referred to as $n$-mode PCA \cite{kapteyn1986approach} and Higher-order SVD \cite{de2000multilinear} for its correspondence to two-way PCA and SVD. The fields of application of the Tucker decomposition are numerous. To mention some, it is applied to signal processing  \cite{de2004dimensionality}, to problems in psychometrics \cite{kiers2001three} and chemical analysis \cite{henrion1994n}. Other applications in computer vision, image recognition, text analysis and optimization are also well documented in the literature \cite{kolda2009tensor,kroonenberg1983three, cichocki2009nonnegative}.
	For a $3$-rd order tensor $\T{X} \in \R^{I \times J \times K }$, the decomposition involving $P$, $Q$ and $R$ components in the first, second and in the third mode perspectively, takes the form of the \textit{n-mode} product (defined in equation \ref{n_mode_prod} of appendix \ref{optens}), i.e.:
	
	 \begin{equation}\label{t}
	 \T{X} \approx \T{g} \times_1 A \times_2 B \times_3 C, \end{equation}
	 
	where $\T{g}\in \R^{P \times Q \times R}$ is a core tensor containing weights of the relationship between the different dimensions of the tensor  $\T{X}$ and $A \in \R^{P \times I},$ $B\in \R^{Q \times J}$ and $C\in \R^{R \times K}$ are the factor matrices for each dimension. A graphical representation of the Tucker model for a $3$-rd order tensor is shown in figure \ref{fig:tucker} below.
	
	\begin{figure}[H]
		\centering
		\begin{tikzpicture}[scale=.75,namenode/.style={scale=.85}]

		\def\ix{3} %
		\def\iy{3} %
		\def\iz{2.5} %

		\def\corescale{1.75}
		\def\rx{\ix/\corescale}
		\def\ry{\iy/\corescale}
		\def\rz{\iz/\corescale}

		\coordinate (XFrontLowerLeft) at (0,0);
		\draw (XFrontLowerLeft) rectangle ++ (\ix,\iy); %
		\begin{scope}[shift={(XFrontLowerLeft)},canvas is zx plane at y=\iy,rotate=90]
		\draw (0,0) rectangle ++ (\ix,\iz); %
		\end{scope}
		\begin{scope}[shift={(XFrontLowerLeft)},canvas is zy plane at x=\ix,rotate=90]
		\draw (0,0) rectangle ++ (\iy,\iz); %
		\end{scope}
		\node[namenode] at ($(XFrontLowerLeft) + (0.5*\ix, 0.5*\iy)$)  {\huge{$\T{X}$}};

		\coordinate (ApproxCtr) at ($(XFrontLowerLeft) + (\ix+0.4*\iz,0.75*\iy) + (0.75,0)$);
		\node[namenode] at (ApproxCtr) {$\approx$};

		\coordinate (U1LowerLeft) at ($(ApproxCtr) - (0,0.75*\iy) + (0.75,0.5)$);
		\draw (U1LowerLeft) rectangle ++ (\ry,\iy);
		\node[namenode] at ($(U1LowerLeft)+(0.5*\ry, 0.5*\iy)$)  {$A$};

		\coordinate (GFrontLowerLeft) at ($(U1LowerLeft) + (\ry+0.5,1)$);
		\draw (GFrontLowerLeft) rectangle ++ (\rx,\ry);
		\begin{scope}[shift={(GFrontLowerLeft)},canvas is zx plane at y=\ry,rotate=90]
		\draw (0,0) rectangle ++ (\rx,\rz);
		\end{scope}
		\begin{scope}[shift={(GFrontLowerLeft)},canvas is zy plane at x=\rx,rotate=90]
		\draw (0,0) rectangle ++ (\ry,\rz);
		\end{scope}
		\node[namenode] at ($(GFrontLowerLeft)+(0.5*\rx,.5*\ry)$)  {$\T{G}$};

		\coordinate (U2LowerLeft) at ($(GFrontLowerLeft) + (\rx+\rz*0.4+0.5,0.5)$);
		\draw (U2LowerLeft) rectangle ++ (\ix,\rx); %
		\node[namenode] at ($(U2LowerLeft)+(0.5*\ix,0.5*\rx)$)  {$B$};
		
		\coordinate (U3LowerLeft) at ($(GFrontLowerLeft) + (0.6,\ry+.8)$);
		\begin{scope}[shift={(U3LowerLeft)},canvas is zx plane at y=0,rotate=90]
		\draw (0,0) rectangle ++ (\rz+0.3,\iz+0.1); %
		\end{scope}
		\node[namenode] at ($(U3LowerLeft)+(1.3,0.5)$) {$C$};

		\end{tikzpicture}
		\caption{Tucker decomposition of a $3$-rd order tensor.}
		\label{fig:tucker}
	\end{figure}
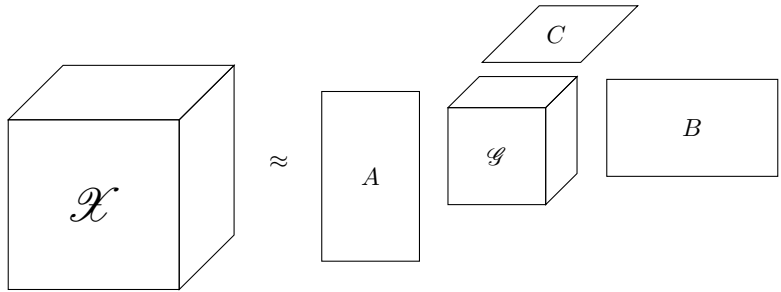							
	
	equation \ref{t} can be rewritten in terms of the outer product as:
	
		\begin{equation} \label{t_op}
	\T{X} \approx \sum_{p=1}^{P}\sum_{q=1}^{Q}\sum_{r=1}^{R}\textbf{g}_{pqr} \left( \textbf{a}_p \circ \textbf{b}_q \circ \textbf{c}_r \right),
	\end{equation}
	where $\textbf{g}_{pqr}$ is the a core element and $\textbf{a}_p$, $\textbf{b}_q$ and  $\textbf{c}_r$ are vector components.
	Equation \ref{t_op} makes the comparison between the PARAFAC and Tucker models easier. It is possible to see that if $P=Q=R$ and the core tensor $\T{G}$ is superdiagonal, i.e. $g_{pqr}\neq 0$ iff $p=q=r$, the two models are equivalent. This means that the PARAFAC can be seen as a special (constrained) case of the Tucker model. However, the higher flexibility embedded in the Tucker model comes at a cost. The Tucker model is generally less parsimonious than the PARAFAC model, unless we are dealing with a strongly dimension-asymmetric tensor. However, it is not generally the case since in addition to the PARAFAC model, the Tucker model has the core tensor to be estimated, which can be high-dimensional itself. Another issue concerning the Tucker model is the lack of uniqueness. It is in fact possible to modify the core tensor by some nonsingular matrices and apply the inverse of those matrices to the factor matrices without altering the fit of the model. The Tucker model is then said to be unique up to rotation of the components. As in the case of PARAFAC, the factors represent the information of each dimension of the original tensor in the latent space. 
		\subsection{Slice-diagonal tensor decomposition}\label{sdt_sec}
\begin{figure}[H]
		\centering
		\begin{tikzpicture}[scale=.75,namenode/.style={scale=.85}]

		\def\ix{3} %
		\def\iy{3} %
		\def\iz{2.5} %

		\def\corescale{1.75}
		\def\rx{\ix/\corescale}
		\def\ry{\iy/\corescale}
		\def\rz{\iz/\corescale}

		\coordinate (XFrontLowerLeft) at (0,0);
		\draw (XFrontLowerLeft) rectangle ++ (\ix,\iy); %
		\begin{scope}[shift={(XFrontLowerLeft)},canvas is zx plane at y=\iy,rotate=90]
		\draw (0,0) rectangle ++ (\ix,\iz); %
		\end{scope}
		\begin{scope}[shift={(XFrontLowerLeft)},canvas is zy plane at x=\ix,rotate=90]
		\draw (0,0) rectangle ++ (\iy,\iz); %
		\end{scope}
		\node[namenode] at ($(XFrontLowerLeft) + (0.5*\ix, 0.5*\iy)$)  {\huge{$\T{X}$}};

		\coordinate (ApproxCtr) at ($(XFrontLowerLeft) + (\ix+0.4*\iz,0.75*\iy) + (0.75,0)$);
		\node[namenode] at (ApproxCtr) {$\approx$};

		\coordinate (U1LowerLeft) at ($(ApproxCtr) - (0,0.75*\iy) + (0.75,0.5)$);
		\draw (U1LowerLeft) rectangle ++ (\ry,\iy);
		\node[namenode] at ($(U1LowerLeft)+(0.5*\ry, 0.5*\iy)$)  {$A$};

		\coordinate (GFrontLowerLeft) at ($(U1LowerLeft) + (\ry+0.5,1)$);
		\draw (GFrontLowerLeft) rectangle ++ (\rx,\ry);
		\begin{scope}[shift={(GFrontLowerLeft)},canvas is zx plane at y=\ry,rotate=90]
		\draw (0,0) rectangle ++ (\rx,\rz);
		\end{scope}
		\begin{scope}[shift={(GFrontLowerLeft)},canvas is zy plane at x=\rx,rotate=90]
		\draw (0,0) rectangle ++ (\ry,\rz);
		\end{scope}
		
		\node[namenode] at ($(GFrontLowerLeft)+(0.2*\rx,.8*\ry)$)  {$\lambda$};
		\node[namenode] at ($(GFrontLowerLeft)+(0.8*\rx,.2*\ry)$)  {$\lambda$};
		\node[namenode] at ($(GFrontLowerLeft)+(0.2*\rx,.2*\ry)$)  {$0$};
		\node[namenode] at ($(GFrontLowerLeft)+(0.8*\rx,.8*\ry)$)  {$0$};
		\node[namenode] at ($(GFrontLowerLeft)+(1.15*\rx,.45*\ry, .0*\rz)$)  {\textcolor{gray}{$\lambda$}};
		\node[namenode] at ($(GFrontLowerLeft)+(0.45*\rx,.45*\ry, .0*\rz)$)  {\textcolor{gray}{$0$}};
		\node[namenode] at ($(GFrontLowerLeft)+(1.4*\rx,1.4*\ry, .8*\rz)$)  {\textcolor{gray}{$0$}};
		\node[namenode] at ($(GFrontLowerLeft)+(0.4*\rx,1.15*\ry, .0*\rz)$)  {\textcolor{gray}{$\lambda$}};
		

		\coordinate (U2LowerLeft) at ($(GFrontLowerLeft) + (\rx+\rz*0.4+0.5,0.5)$);
		\draw (U2LowerLeft) rectangle ++ (\ix,\rx); %
		\node[namenode] at ($(U2LowerLeft)+(0.5*\ix,0.5*\rx)$)  {$B$};
		
		\coordinate (U3LowerLeft) at ($(GFrontLowerLeft) + (0.6,\ry+.8)$);
		\begin{scope}[shift={(U3LowerLeft)},canvas is zx plane at y=0,rotate=90]
		\draw (0,0) rectangle ++ (\rz+0.3,\iz+0.1); %
		\end{scope}
		\node[namenode] at ($(U3LowerLeft)+(1.3,0.5)$) {$C$};

		\end{tikzpicture}
		\caption{SDT decomposition of a $3$-rd order tensor.}
		\label{fig:sdt}
	\end{figure}
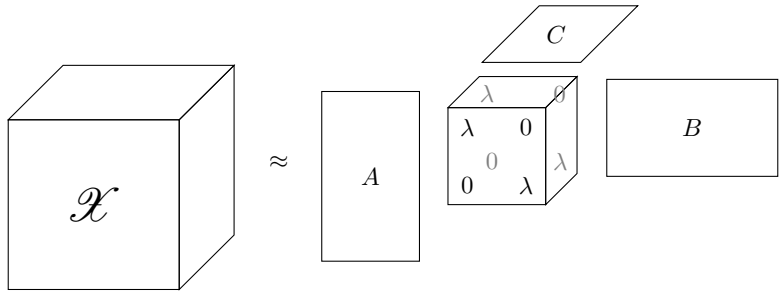

After introducing the two main tensor factorization methods, let us introduce here a new tensor decomposition method which represents a more flexible version of the PARAFAC model. Similarly to the Tucker decomposition, this methodology allows to impose different dimensions over each mode factors. Nevertheless, it differs from the Tucker version since it is characterized by a core tensor which is slice-diagonal. This means that the slices\footnote{Definition of tensors slices can be found in appendix \ref{optens}.} of the core tensor are diagonal matrices. Depending on the structure of the tensor and the application at hand, one would impose the diagonal constraint to the horizontal, lateral or frontal slice. For the application reported in this paper, we use frontal diagonal slices. This factorization, which we name Slice-diagonal tensor (SDT) decomposition, is more parsimonious and it embeds stronger uniqueness properties with respect to the Tucker model given less rotational freedom of the components. Indeed, the mathematical representation is equivalent to the Tucker decomposition, i.e.:

	 \begin{equation}\label{sd}
	\T{X} \approx \T{\varLambda} \times_1 A \times_2 B \times_3 C, \end{equation}
	where $\T{\varLambda}\in \R^{P \times Q \times R}$ is a slice-diagonal core tensor with elements $\lambda_{pqr}\neq0$ iff $p=q$ and $A \in \R^{P \times I},$ $B\in \R^{Q \times J}$ and $C\in \R^{R \times K}$ are the factor matrices. It is easy to see that when the third mode collapses to $1$ and $P=Q$, this is equivalent to the SVD. A graphical representation of the SDT decomposition of a $3$-rd order tensor is represented in figure \ref{fig:sdt}.
	
	This new decomposition technique has several advantages. The first one is surely the parsimony of the model, which does not eliminate the flexibility of the Tucker model. To better understand this feature, it is necessary to analyse the number of elements to be estimated for each model in the case of a $3$-rd order tensor. We  analyse the model size related to PARAFAC, Tucker and SDT and rely on the notation used in equations \ref{p}, \ref{t} and \ref{sd} for the dimensionality of the components.  Table \ref{tab1} below shows the model complexity of the three decompositions:
	
	\begin{table}[H]
		\centering
		
		\begin{tabular}{|l|l|}
			\hline
			\textbf{Decomposition model} & \textbf{Number of free parameters} \\ \hline
			PARAFAC             & $R \times (I+J+K)$             \\ \hline
			Tucker              & $(P \times I) + (Q \times J) + (R \times K)+ (P \times Q \times R)$    \\ \hline
			SDT                 & $(P \times I) + (Q \times J) + (R \times K)+ (P \times R)$        \\ \hline
		\end{tabular}
	\caption{Number of free parameters to be estimated in each model}
	\label{tab1}
	\end{table}
	
	It is clear that the Tucker model, with exception of some rare circumstances, is the model with the highest number of elements to be estimated. Nevertheless, the flexibility given by the core tensor of this model increases the possibility to achieve a decomposition, which is not granted in the PARAFAC. In contrast, the SDT is more parsimonious than the Tucker decomposition since, as it is possible to notice from the table, the core tensor has only $(P \times R)$  terms instead of $(P \times Q \times R)$. The difference rapidly increases as $Q$ increases. 	
	A second advantage is that the restrictions on the core tensor reduce the possible rotation of the components, improving uniqueness of the decomposition. A final advantage of this representation is achieved when we deal with skewed, asymmetric tensors, e.g. the bilateral trade matrix. In this context, the Tucker decomposition is unable to retrieve the full asymmetry in the bilateral exposures. This is because part of this asymmetry is incorporated on the core tensor off-diagonal values, while the remaining part is incorporated in the factor matrices. On this issue, \cite{nickel2011three} proposed the RESCAL decomposition which restricts the Tucker-2 model\footnote{Tucker-2 decomposition is a Tucker factorization with only two factor matrices estimated and the remaining constrained to be the Identity matrix.} such that the factor matrices over the bilateral exposures are constrained to be equal so that the asymmetry is only reflected in the core tensor. However, the RESCAL considers the decomposition in a core tensor and the two factor matrices, while in our context we want to have the third, time dependent factor matrix. In contrast to RESCAL, the SDT decomposition, having a core tensor which is slice-diagonal, has only the variance weights on the core and any asymmetry in the data is embedded in the factor matrices. The factor components of the SDT decomposition, as for the other ones, are estimated via an optimization problem which minimizes the Frobenious norm of the error with respect to the data tensor $\T{X}$ \cite{kolda2009tensor}. This optimization is a non-linear least square problem that cannot be solved with a simple operation. To solve this optimization, we rely on an iterative algorithm such as the Alternating least squares (ALS). ALS was introduced for PARAFAC by \cite{carroll1970analysis, harshman1970foundations} and for the Tucker decomposition by \cite{kroonenberg1980principal, kapteyn1986approach}. This methodology solves the optimization problem by dividing it into small least squares problems. This allows to solve the optimization for each factor separately, keeping the other ones constant at the previous iteration values. Then, it iterates alternating among all the factors until the algorithm converges or a maximum number of iterations is reached. The ALS has several applications and it is workhorse algorithm for Tensor factorization.\footnote{A pseudocode of this procedure can be found in appendix \ref{als_algo}.}

	\subsection{Model selection}\label{MS}
	Before discussing the application of the three different factorizations, a key topic that has to be discussed first is the choice of the number of factors to be used in the decompositions. In some circumstances, the number of components comes from a theoretical analysis, e.g. in Chemometrics. In absence of theoretical guidelines we rely on data-driven approaches for model selection. We use an Information Criterion (IC) to deduce the best model specification to use for each factorization. In particular, we mention among the information criteria, the Bayesian information criterion ($BIC$) \cite{schwarz1978}, the Akaike information criterion ($AIC$)\cite{akaike1974} and the corrected (for finite samples) Akaike information criterion ($AIC_C$)\cite{CAVANAUGH1997}. The $BIC$, $AIC$ and $AIC_C$ are computed as follows:
	
	\begin{equation}\label{eq_BIC}
	BIC= u\ln\left( \frac{SSR}{u}\right) +w\ln(u)	
	\end{equation}	
	
	\begin{equation}\label{eq_AIC}
	AIC= u\ln\left( \frac{SSR}{u}\right)+2w
	\end{equation}
	
	\begin{equation}\label{eq_AICC}
	AIC_C=u\ln\left( \frac{SSR}{u}\right) + 2w\left(\frac{u}{u-w-1}\right)
	\end{equation}
	
	where $u$ is the number of data-points, $w$ is the number of estimated elements and $SSR$ are the sum of squared residual of the factorization. The $AIC$ is the one, among the information criteria discussed, which penalizes less non-parsimonious models while its correction, the $AIC_C$, penalizes much more over-parametrized models in case of small sample data. Finally, the $BIC$ penalizes more for the number of estimated elements compared to the Akaike information criteria. Other criteria have been proposed in the tensor decomposition literature. Among all, the DIFFIT criterion of \cite{diffit} and the CONsistency CORe DIAgnostic (CONCORDIA or CORCONDIA) \cite{concordia} for the PARAFAC decomposition are often used. In particular, the DIFFIT criterion computes the increase of the fit (the difference in fitting) for models with increasing total number of components.\footnote{For models with the same complexity, e.g. a Tucker decomposition with components dimension {2,1,3} has the same total number of components as the {2,2,2} specification, only the one with maximum fit is taken into account.} The one for which the relative increase with respect to the previous model specification is maximized is the chosen model. However, being an heuristic method to chose the number of components, for datasets for which few components are able to reproduce a great portion of the variability, the method severely under-parametrizes the factorizations. On the other end, CONCORDIA is a well suited criterion used for PARAFAC decomposition since this model is not nested, hence a model with $R$ factors is not equal a model with $R-1$ factors plus 1 additional component. Therefore, without imposing any constraints, e.g. orthogonality over all the dimensions, the standard information criteria are not well suited. The idea behind of the CONCORDIA criterion comes from the relationship between PARAFAC and Tucker decomposition we mentioned in equation \ref{t_op} for which the PARAFAC can be seen as a constrained Tucker with a superdiagonal core tensor. The criterion is computed as:

	\begin{equation}\label{eq_CONCORDIA}
	CONCORDIA=100\left(1- \frac{\sum_{p=1}^{P}\sum_{q=1}^{Q}\sum_{r=1}^{R}(\textbf{g}_{pqr}-\textbf{d}_{pqr})^2}{\sum_{p=1}^{P}\sum_{q=1}^{Q}\sum_{r=1}^{R}\textbf{d}_{pqr}^2}\right),
	\end{equation}
	
where $\textbf{d}_{pqr}$ is a super-diagonal tensor. If the core tensor $\textbf{g}_{pqr}$ is effectively superdiagonal, the CONCORDIA criterion is near to $100$ while if model is over-parametrized or the PARAFAC is not the correct structure, it falls toward $0$. The general practise is to take the higher number of components for which the CONCORDIA value is high (between $80$ and $100$) \cite{Sena2006}. For the reasons aforementioned, we use the CONCORDIA criterion for the PARAFAC decomposition and the $BIC$ for Tucker and SDT decompositions.

\subsection{Hidden Correlation matrix}\label{HCM_build}
After the decomposition is carried out, we build the hidden correlation matrix (HCM). The multilinear decomposition acts as a dynamic whitening of the correlation from market fluctuations. This procedure cleans the factor loading from any time specific behaviour, leaving only the long-run features of the links. Taking the decomposition induced by equation \ref{sd}, we construct a matrix, which we call link matrix  ($\Gamma$), representing the dependencies between stocks. The link matrix $\Gamma$ is computed as:
\begin{equation}\Gamma=A\tilde{\Lambda}B'\end{equation} 

where $A \in \R^{P \times I}$ and  $B\in \R^{Q \times J}$ are the cross-sectional factor matrices and $\tilde{\Lambda} \in \R^{P \times Q}$ is the matricization\footnote{Matricization definition and computation can be found in appendix \ref{optens2}.} of the core tensor $\Lambda$ of the SDT decomposition.\footnote{One would use the $G$ core tensor if the Tucker decomposition is employed and an identity matrix of size $R \times R$ in case the PARAFAC is used.} In the specific case of correlation matrices (or other symmetric matrices), $P=Q$ and $I=J=M$, where $M$ is the number of stocks. The corresponding hidden correlation matrix $\Omega \in \R^{M \times M}$ is defined as the normalized version of $\Gamma$, in the same way of a covariance matrix,\footnote{See appendix \ref{Corr_sec} for correlation matrix computation procedure in scalar and matrix form.} i.e.: 
	
	\begin{equation}
	\Omega=D^{-1} \Gamma D^{-1}
	\end{equation} 
	
	where $D$ is a diagonal matrix defined as $D_{mm}=\sqrt{\Gamma_{mm}}$ for $m=1,\dots , M$. After the normalization, we have a a matrix with ones on the main diagonal and bounded values on the off-diagonal. However, to define a matrix as \textit{correlation matrix}, it is not enough to have a symmetric matrix with bounded values between $-1$ and $+1$ and with all $1s$ on the main diagonal. A correlation matrix must fulfil an additional property, which is to be positive semi-definite. In this regard, \cite{higham2002computing} introduced a way to compute the closest correlation matrix given a symmetric matrix $W$ using a weighted Frobenius norm loss function and an alternating projection method. The optimization problem writes:
	
	\begin{equation} \argmin_Z \{\lVert W- Z\lVert_F: Z \text{ is a correlation matrix}\}\end{equation}
	where $W$ is the sample correlation matrix and $Z$ is the closest correlation matrix with respect to the Frobenious norm $\lVert \cdot \lVert_F$. In practical terms, the methodology minimizes the distance between $W$ and a generated correlation matrix $Z$, where $Z$ fulfils all the requirements of a correlation matrix.
	We refer to \cite{higham2002computing} for technical details on the numerical method implemented in the optimization procedure.
	To prevent the hidden correlation matrix to be ill posed, we compute the closest correlation matrix $\Theta$ from $\Omega$, i.e.:
	\begin{equation}\label{ccm}
	\argmin_\Theta\{\parallel\Omega-\Theta  \parallel_F:\Theta \text{ s a correlation matrix}\}
	\end{equation}
		and $\Theta$ will be the correlation matrix cleaned by time specific fluctuations. 
		As explained in this section, the hidden correlation matrix is retrieved by following a precise methodology, which is reported in the schematic diagram below:\\

		\begin{center}
			\cascadingblocks{{\textbf{Input:} $3$-rd order Covariance tensor of size $M \times M \times T$},{Tensor Factorization: SDT, Tucker or PARAFAC},{Model selection using Information Criteria: BIC, AIC, DIFFIT or CONCORDIA},{Combination of factor components and normalization},{\textbf{Output:} Hidden (low-rank) correlation matrix of size $M \times M$}}
		\end{center}	
		
		Having delineated the methodology's pipeline, let us now run a simulation to test the full procedure.
		\section{Simulation study}\label{sim_sec}

	In this section we numerically test the HCM construction procedure presented in the previous section. In particular, we show the results for the SDT decomposition and then compare it to Tucker and PARAFAC. To this end, we generate a Covariance tensor for which we know the structural correlation matrix and the time component. First of all, we compose a block-diagonal correlation matrix by generating blocks (of different size) from the Vine-Beta (VB) model proposed in \cite{LKJ2009}. The method samples partial correlations from a Beta distribution and then converts them to the correlation by using a recursive formula. A parameter $d$ can be used to tailor the strength of the correlations with small values of $d$ generating correlation near the $[-1,1]$ boundaries. Compared to other correlation matrix simulations, the VB model permits to have very strong correlation structures. We proceed as follows:
	
	\begin{enumerate}
	\item Generate $5$ blocks of dimensions 20, 10, 30, 15 and 25 using the VB model with $d=0.2$; 
	\item Store the blocks on the diagonal of a $100 \times 100$ empty matrix that we name $E$;
	\item Generate a $100 \times 100$ correlation matrix $S$ using the VB model with $d=1$;
	\item Form a new matrix $\Omega$ combining the previous two matrices as $\Omega=0.9E+0.1S$;
	\item Generate a vector of variances $v$ and form the Covariance matrix $\Sigma=\left(diag(\sqrt{v})\right)\Omega\left(diag(\sqrt{v})\right)$;
	\item Compute the SVD of $\Sigma$ and rewrite it with $10$ factors resulting in  $\Sigma_{SVD}$;
	\item Create a tensor $\T{X}$ using the n-mode product between $\Sigma_{SVD}$ and a time component. 
	\end{enumerate} 
	
	Points $1$-$2$ generate the block structure of the final correlation matrix while point $3$ is used to fill the other entries of the correlation matrix $\Omega$.\footnote{In point 4 we used the combination $0.9$-$0.1$ to build a matrix $\Omega$ having a strong block-diagonal structure and use the closest correlation algorithm \cite{higham2002computing} on $\Omega$ to ensure it fulfils all the requirement of a correlation matrix.} Points $5$-$6$ build the covariance matrix $\Sigma$ and its rank-reduced form using the SVD components. Finally, we generate the tensor $\T{X}$ as the n-mode product between $\Sigma_{SVD}$ and a time component (the VIX time series in our case) and we add to it a standard Gaussian noise tensor $\T{V}$ of the same size.\footnote{We have also run the same simulation changing the size and number of blocks, the parameter $d$ and the number of factors in the SVD in point $6$. The results are equivalent to the ones reported.}
	 $\T{X}$ is a $3$-rd order tensor with dimensions $100 \times 100 \times 150$ which has $10$ correlation factor components. We now proceed in running the HCM building methodology based on the SDT decomposition in order to asses the capabilities of such technique in retrieving the true correlation structure and time component. 
	\subsection{Simulations results}
\begin{figure}[H]		
	\centering	
	\includegraphics[width=0.9\linewidth,height=0.25\textheight]{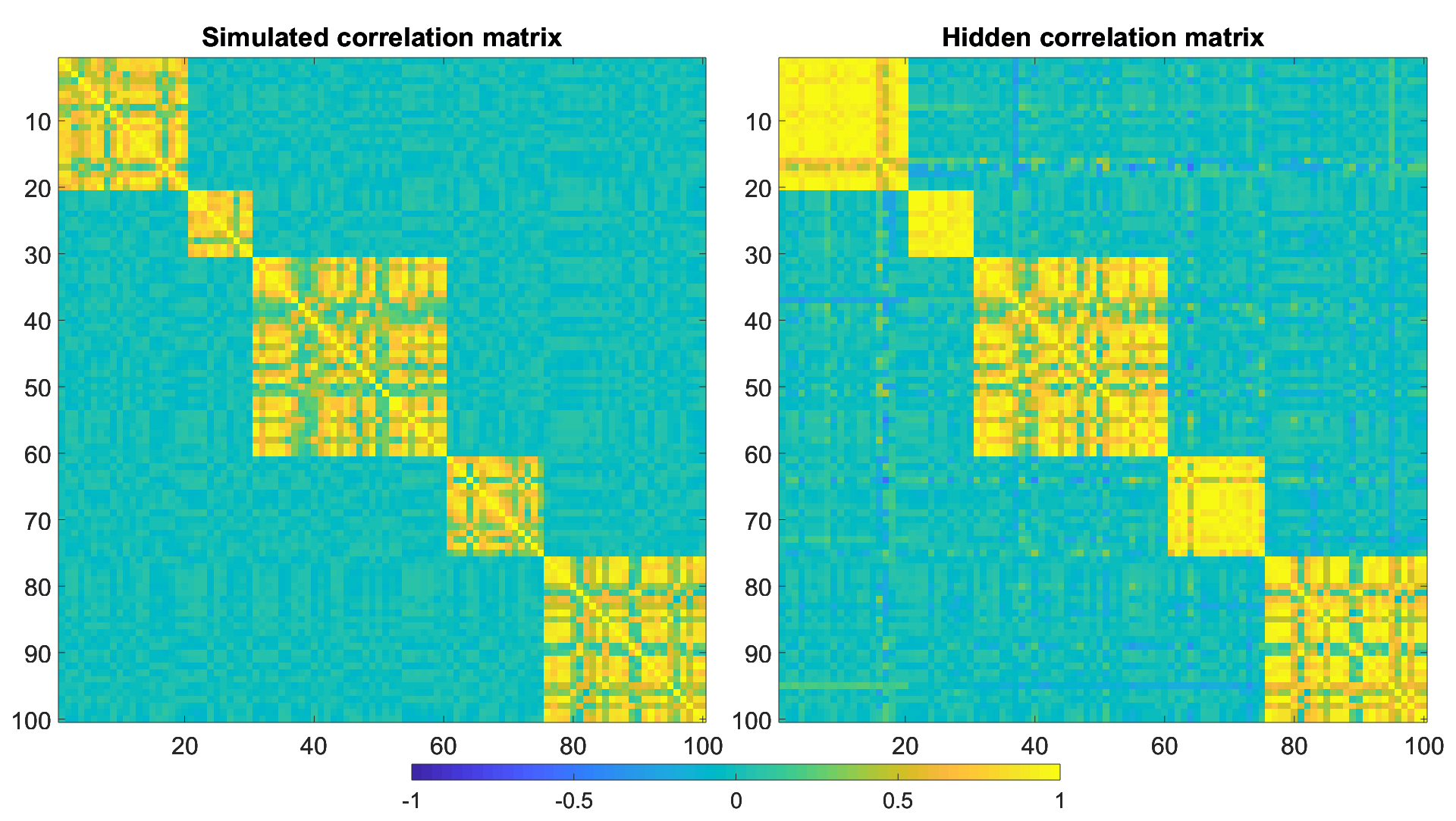}	
	\caption{Comparison between the simulated correlation matrix $\Omega$ used to construct the tensor Hidden correlation matrix. Colours go from blue to yellow in the range [-1, 1].}
	\label{fig:sim_corr}		
\end{figure}

	Before evaluating the goodness of the decomposition, we need to establish which is the number of components to use for the SDT decomposition. The BIC shows a minimum value in correspondence of $10$ correlation components, suggesting that the decomposition is able to correctly identify the number of components we used to construct the tensor.\footnote{Figure \ref{fig:sim_BIC} in appendix \ref{sim_res_plots} depicts the BIC for different numbers of correlation components.} After the optimal number of components to use have been found, the SDT decomposition is carried over and this results in $4$ components, i.e. the core tensor, the two cross-sectional factor matrices and the time dependent component. Regarding the latter, we report that it matches almost perfectly the VIX time series.\footnote{Figure \ref{fig:sim_time} in appendix \ref{sim_res_plots} shows the results of the comparison.} Finally, comparing the simulated correlation matrix $C$ with the hidden correlation matrix computed as in equation \ref{ccm}, we can see from figure \ref{fig:sim_corr} that the whole structure is recovered and the magnitude of the correlation preserved. This result shows that the tensor factorization can retrieve the structural (hidden) matrix. It is important to recall that, this methodology is based on the assumption that such a matrix exists and is stable over time. In fact, if the correlation matrix undergoes a structural change for which the correlations are permanently changed, this kind of procedure would not be optimal even if still preferable to the classic correlation matrix.

In order to test if the procedure is able to unveil the true correlation even in different samples, we need to analyse the methodology against different sample periods. To achieve this, we now proceed to apply the HCM procedure to two non-overlapping samples in order test if the ability to retrieve the correlation structure doesn't change. We then report the same kind of results for the PARAFAC and Tucker decompositions for benchmark analysis.

	\subsection*{Time dependency analysis}
	To test if the structural (hidden) correlation matrix can be retrieved with only a part of the sample, we subdivide the covariance tensor in two smaller tensors originating from non-overlapping samples. In particular, we split the tensor in the middle of the third dimension (time dimension), generating two covariance tensors of size $100 \times 100 \times 75$, which . In principle, we should retrieve the same correlation matrix since it is the one used to generate the full tensor. The computation of the BIC results in a minimum at $10$ correlation components in both the samples, which is the number used to build then full covariance tensor.\footnote{We report the results in figure \ref{fig:sim_BIC2} in appendix \ref{sim_res_plots}.} In figure \ref{fig:sim_HCM2} we show the results of the two HCMs built over the two non-overlapping samples. We can easily see that the simulated correlation structure is retrieved in both the sample periods and that the HCMs are almost identical between themselves and the simulated correlation matrix. This results shows that if the assumption of the existence of the structural correlation matrix is fulfilled, the methodology is able to unveil it. For this reasons, the hidden correlation matrix is a feasible alternative to the standard correlation matrix. In particular, this matrix could be used to test risk management model and build portfolios. 

	\begin{figure}[H]		
	\centering	
	\includegraphics[width=1\linewidth,height=0.25\textheight]{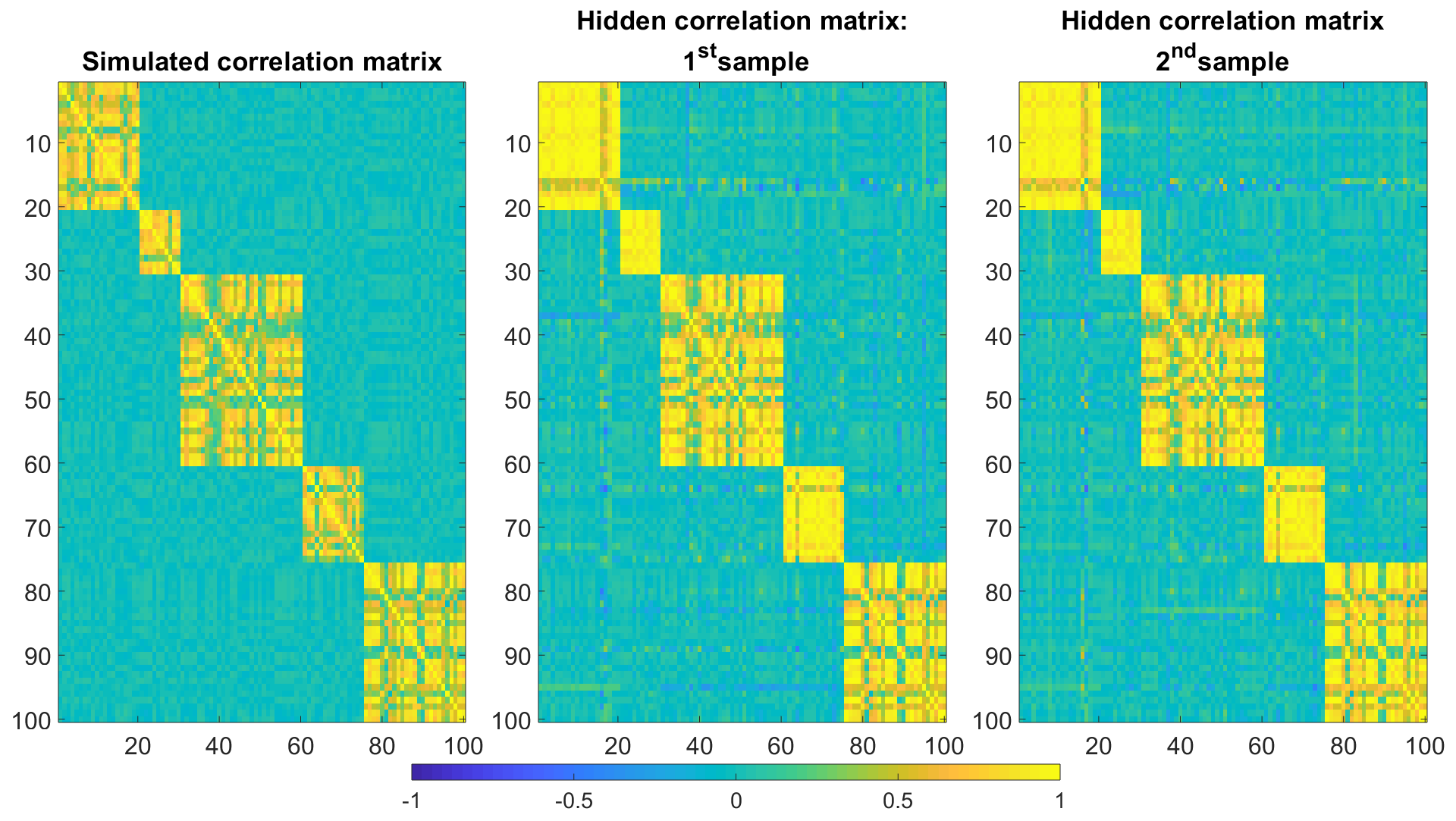}		
	\caption{Comparison between the simulated correlation matrix $\Omega$ used to construct the tensor and the hidden correlation matrix built over the two non-overlapping samples using the SDT decomposition. Colours go from blue to yellow in the range [-1, 1].}
	\label{fig:sim_HCM2}	
\end{figure}
In the following subsection we compare the results just shown with the ones got using different decomposition techniques.
		\subsection*{Model comparison}
	To make a benchmark analysis, we report here the results for the Tucker and PARAFAC decomposition approaches. Regarding the Tucker decomposition, the results are almost identical to the SDT. More precisely, the BIC is optimized at $10$ components in both the sample periods as for the SDT.\footnote{Figure related to the BIC for the Tucker model is shown in appendix \ref{sim_res_plots} in figure \ref{fig:sim_HCM22}.}
	\begin{figure}[H]
		\centering
	\includegraphics[width=1\linewidth,height=0.25\textheight]{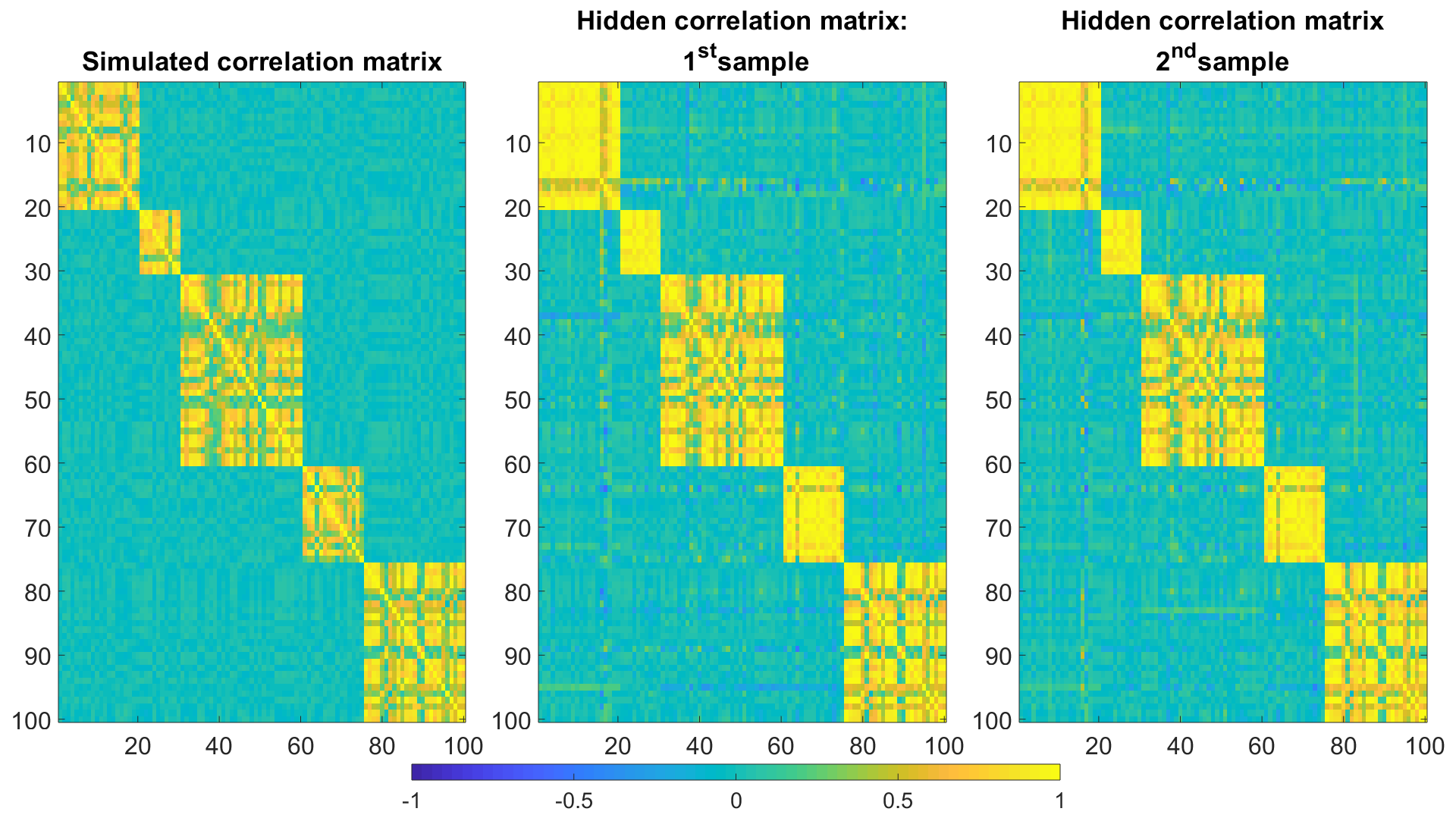}
		\caption{Comparison between the simulated correlation matrix $\Omega$ used to construct the tensor and the hidden correlation matrix built over the two non-overlapping samples using the Tucker decomposition. Colours go from blue to yellow in the range [-1, 1].}\label{sim_tuck_HCM}
	\end{figure}
	 In addition, as it is possible to notice from figure \ref{sim_tuck_HCM}, also the hidden correlation matrices retrieved from this methodology are very similar. This is due to the fact that the Tucker decomposition has almost zero off-diagonal elements in each slice of the core tensor, making it numerically equivalent to the slice-diagonal core tensor of the SDT factorization. Regarding the PARAFAC decomposition, the results are slightly different. First of all, the maximum number of factors for which the decomposition has an admissible CONCORDIA criterion are $6$ for the first factor and $5$ for the second.\footnote{Figure related to the CONCORDIA criterion for the PARAFAC model is shown in appendix \ref{sim_res_plots} in figure \ref{fig:sim_CONCORDIA}.} The number of components selected by the CONCORDIA is almost half the number of true correlation components ($10$) and this is due to the fact that we are imposing the time dimension to have the same number of factors as the correlation dimensions. This makes the PARAFAC model much less robust since in this construction we have only one time factor and the other factors captured by the decomposition are noise. 
	 \begin{figure}[H]
	 \includegraphics[width=1\linewidth,height=0.25\textheight]{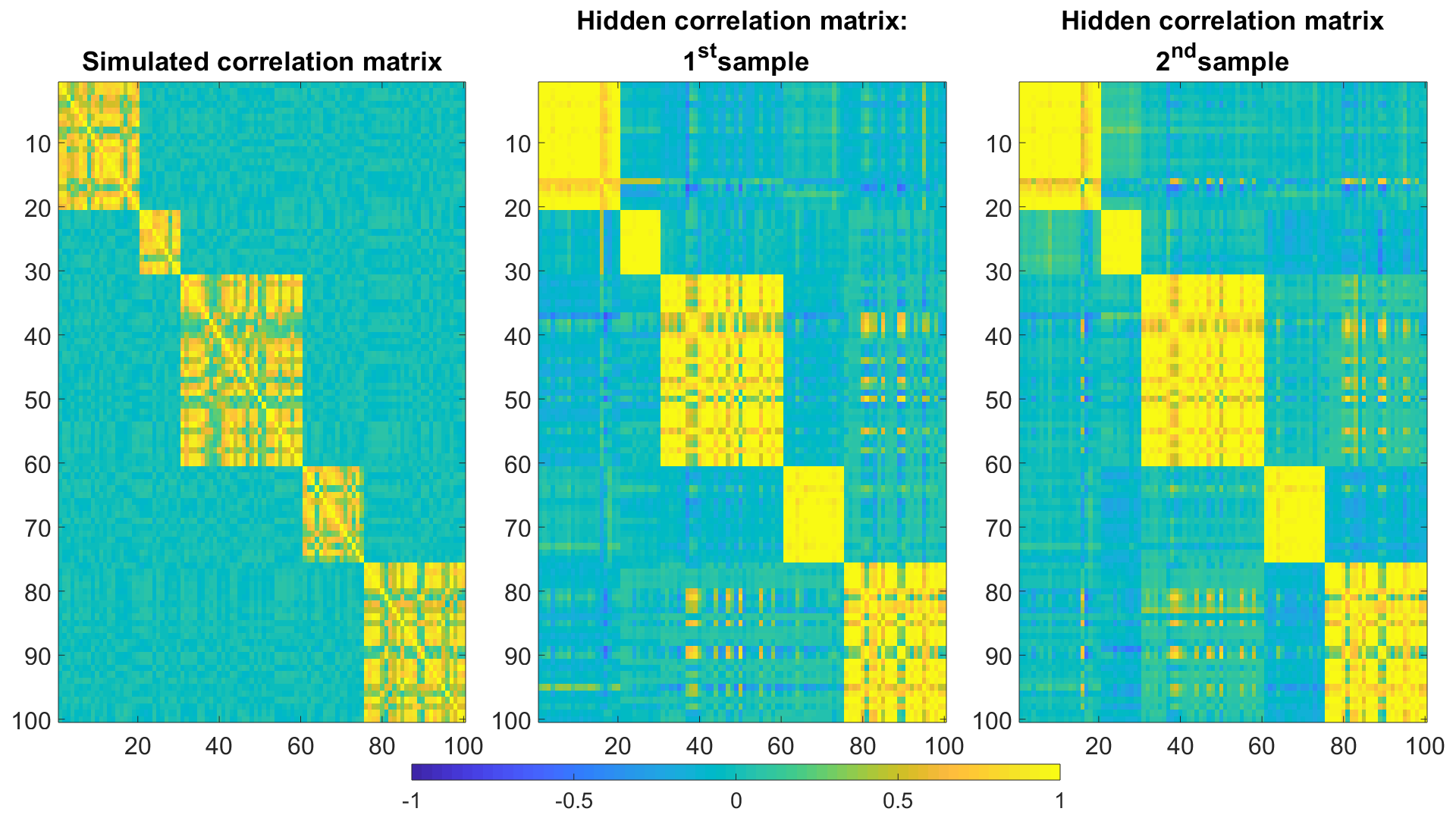}
	 	\caption{Comparison between the simulated correlation matrix $\Omega$ used to construct the tensor and the hidden correlation matrix built over the two non-overlapping samples using the PARAFAC decomposition. Colours go from blue to yellow in the range [-1, 1].}\label{fig:sim_CONCORDIA2}
	 \end{figure}
	 Regarding the hidden correlation matrices, as it is possible to notice from figure \ref{fig:sim_CONCORDIA2}, the structure seems to be a good approximation of the true structure, even if the result is worse than the other two, more flexible, decompositions. This simulation has shown that the procedure, under the assumption of the existence of a structural correlation matrix, is able to unveil it. It further gives us as result that the SDT has better performance with respect to the PARAFAC and identical performance with respect to the Tucker decomposition, even if it is more parsimonious. In the next section we use real data to construct a hidden correlation matrix and analyse its statistical properties.

	\section{Empirical application}\label{app_sec}
	
	In this section we use real financial data to build a covariance tensor from which we extrapolate the hidden correlation matrix following the procedure presented in section \ref{decomp_sec}. We show the results concerning the SDT based procedure and comment on the other factorizations techniques. We firstly present the data used in the analysis and then the retrieved factors used to construct and analyse the hidden correlation matrix. In this case, unlike the simulation example of section \ref{sim_sec}, the correlation structure is not known in advance, so we assume the existence of an hidden correlation matrix which doesn't depend on the sample and we run two non-parametric tests for the statistical equality of the spectrum of the HCM built over two non-overlapping sample periods in order to quantitatively asses this assumption.
	
	\subsection{Data}
	We use a subsample of intra-day stock prices (data is sampled at 5 minutes frequency) of the S\&P100 from January 2005 up to June 2017. In particular, we use $65$ out of the $100$ available stocks since these stocks are observable throughout the entire sample period. With this data, we are able to build the covariance tensor ($\T{X}$), constructed concatenating the covariance matrices over time. We calculate every covariance matrix using one month of data so that we have twelve matrices per year. After the concatenation of all the covariance matrices, we have $3$-rd order covariance tensor of dimension $65 \times 65 \times 150$ ($M \times M \times T$). Each element of the tensor contains the covariances between stocks and at time a specific time occasion.
\subsection{Model selection}
	
\begin{figure}[H]		
	\centering	
	\includegraphics[width=0.8\linewidth,height=0.25\textheight]{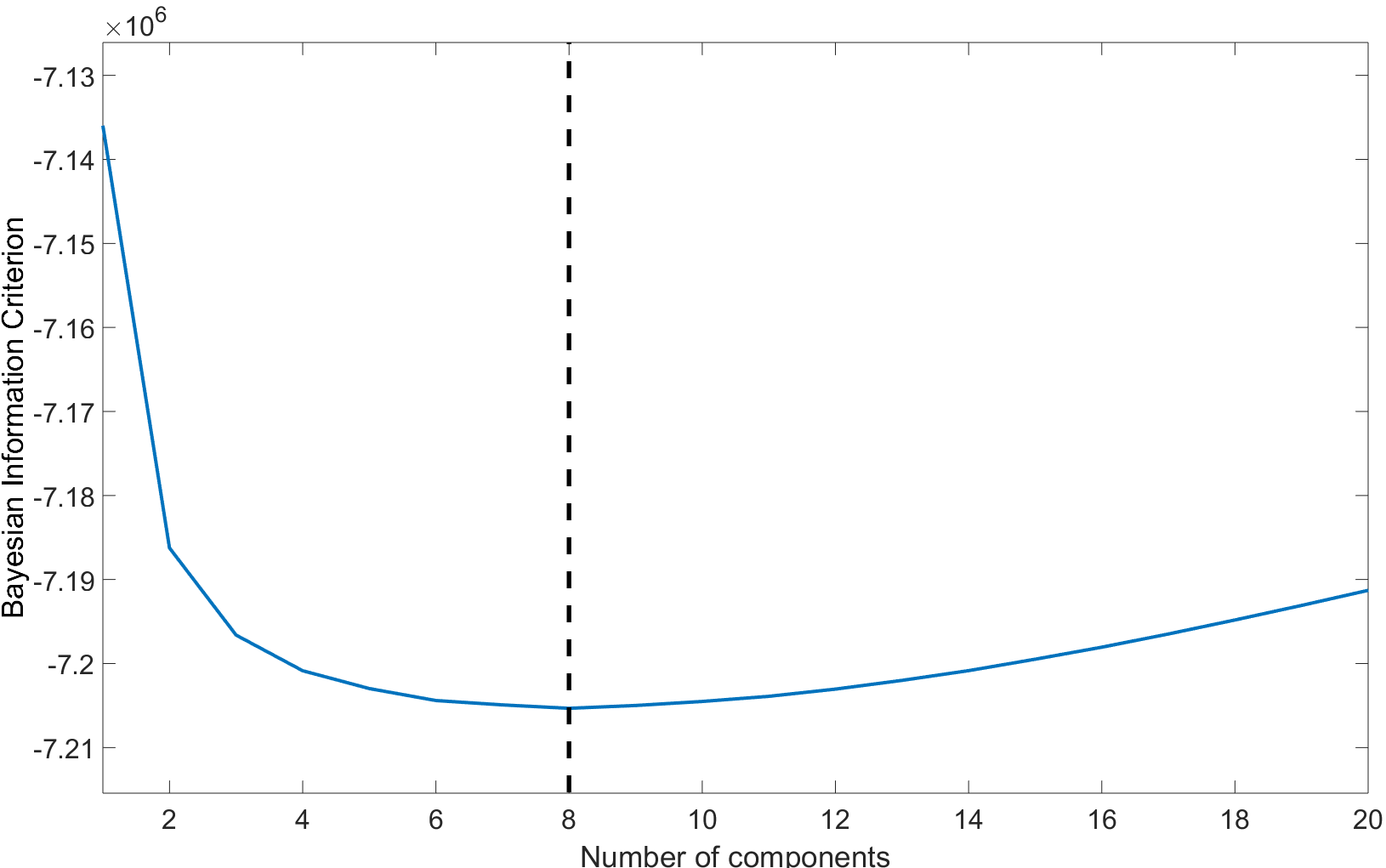}	
	\caption{Bayesian information criterion (BIC) for the number of correlation components of the Tucker and SDT decompositions. The plot shows a minimum for both the model at $8$.}	
	\label{fig:emprirical_IC_SDT}	
\end{figure}

As shown in the previous section, the choice of the model complexity is fundamental. To choose the number of components we rely on the BIC of section \ref{MS} for the SDT and Tucker models and the CONCORDIA criterion for the PARAFAC decomposition. 
Figure \ref{fig:emprirical_IC_SDT} depicts the BIC criterion for the SDT decompositions. As it is possible to notice, the minimum value of the BIC is reached in correspondence to $8$ correlation components.\footnote{A similar plot for the Tucker decomposition can be found in figure \ref{fig:emprirical_IC} in appendix \ref{real_res_plots}. Also for the Tucker decomposition, the optimal number of correlation components selected by the BIC is $8$ .} 
For the PARAFAC decomposition, the CONCORDIA criterion selects a $4$ components specification. The minimum threshold of $80\%$ is reached only for the $2$ and $4$ factors specifications (apart the trivial $1$ factor PARAFAC). As the general procedure requires, we select the $4$ factors specification since it is the one with the higher number of components.\footnote{A plot showing the CONCORDIA for the application is depicted in figure \ref{fig:emprirical_CONCORDIA} of appendix \ref{real_res_plots}.} These results convert to a model complexity of $1198$ for the SDT factorization, $1254$ for the Tucker decomposition and $1120$ for the PARAFAC. This model complexity reduction of the SDT, besides the clear advantage of a lower number of parameters to be estimated, increases the uniqueness properties of the decomposition. Instead, the motivation behind the PARAFAC having this huge number of parameters even for lower number of correlation factors is that it requires to have the same number of factors among all the dimensions, so we have also $4$ time components, while for the Tucker and SDT decomposition we are able to set it to $1$ since we are going to remove the market dynamic, which is one factor.\footnote{It is perfectly possible to generalize the method to $R\geq2$ time factors, even if, in this specific financial application, would become difficult to justify.}	
	
	\subsubsection{Dynamic component}
\begin{figure}[H]			
\centering
	\includegraphics[width=0.9\linewidth,height=0.3\textheight]{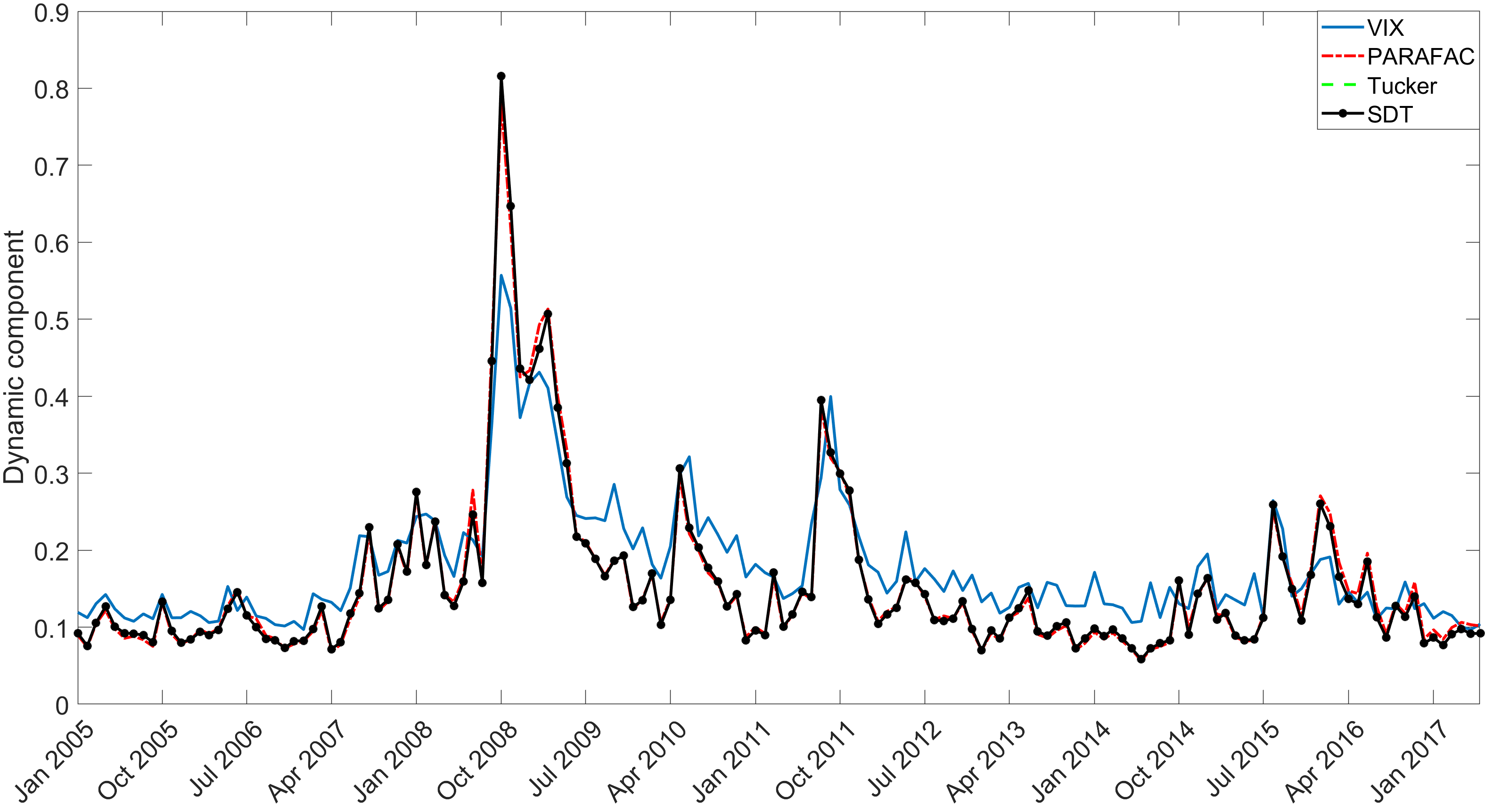}	
		\caption{Comparison between the implied volatility index and the dynamic factor of the different decompositions.}
		\label{fig:vix}	
\end{figure}

	In this subsection we analyse the Dynamic factor estimated with the $3$ different decompositions. To better understand the behaviour of the dynamic components, we compare them with a volatility indicator. In particular, we compare the dynamic component of the covariance tensor with the Volatility index (VIX) of the S\&P. The VIX index is provided by the Chicago Board Options Exchange (CBOE) and is frequently used as volatility benchmark. Figure \ref{fig:vix} depicts the dynamic component\footnote{Since the VIX represents the volatility rather than the variance, we take the square root of the component.} of the three different decomposition techniques.\footnote{For the PARAFAC decomposition we sum the $4$ time components.} We can easily notice that all the decompositions are capable of reproducing the volatility behaviour over time. It is worth to notice that Tucker and SDT are almost overlapping, showing that the lower number of free parameters did not change the fit of the model. There are only few small differences between PARAFAC, Tucker and the SDT decomposition but all of them are able to spot all the main volatility surges and its dynamic. This result empirically shows that the tensor decomposition can be used to disentangle spatial structures from their dynamic component. Furthermore, one could forecast the dynamic component (or use derivatives contracts to extract it) and then recombine it with the HCM to forecast a specific point in time correlation matrix.

	\subsection{Hidden Correlation matrix}
\begin{figure}[H]			
	\centering
		\includegraphics[width=0.9\linewidth,height=0.25\textheight]{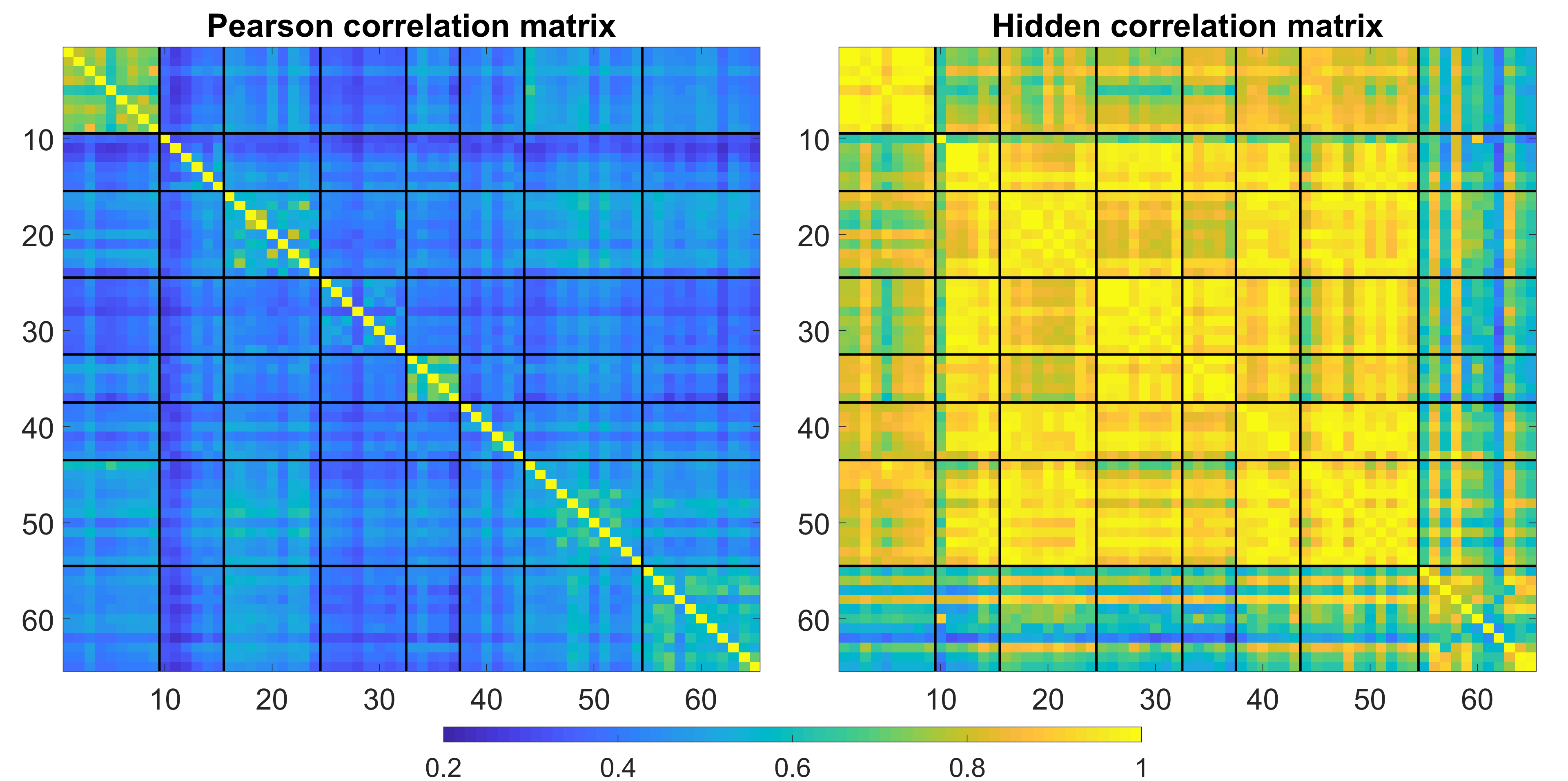}	
			\caption{Comparison between the Pearson correlation matrix and the HCM built from the SDT decomposition. Black lines subdivide the correlation matrix in industry sectors, which are: Basic materials, Consumer goods, Services, Finance, Utilities, Technology, Industrial goods and Healthcare. Colours go from blue to yellow in the range [0.2, 1], where 0.2 and 1 are the lowest and highest values of correlation empirically found.}
			\label{fig:real_data_HCM_full}
	\end{figure}

	The static components represent the stock covariances in the latent space. These factors give information on the linkages between stocks and its magnitude. An high level of each factor loading implies a great effect on the specific component while the same direction is a proxy of the linkages. Stocks which are matched together (both in magnitude and direction) have similar behaviours. The HCM is precisely built on this characteristic of the components. The analysis of such components can unveil very important information regarding the structure of the correlation matrix we construct. In particular, if the first component strongly affects all the industries in the same direction and in an homogeneous way, this means that this factor is common to all the sectors and it is usually attributed to the market component. What we found in our analysis (reported in appendix \ref{stat_comps}) is that the first factor component is homogeneous among all assets across the different decompositions, resulting in the need of a closer analysis when building the hidden correlation matrix, because the market factor can hide the true marginal dependencies between the stocks. For this reason, we firstly show the results with the market mode and then without the market mode. 
	We now analyse the hidden correlation matrix constructed as described in section \ref{HCM_build}. We study two different features of this correlation matrix, i.e spatial and stability properties. Firstly, we study the capabilities of this matrix to cluster the stocks and secondly, whether this feature remains similar in two non-overlapping periods in order to test the decomposition’s ability to retrieve structural features from the data. Finally, we investigate such a relationship through the spectrum of the two HCMs. In particular, we study the distribution of the eigenvalues of the hidden correlation matrices of the two subsamples and perform a Kruskal-Wallis and Kolmogorov-Smirnov non-parametric tests \cite{daniel1990applied} to infer whether the two sets of eigenvalues come from the same distribution. If the two matrices have similar cluster structures and the tests fail to reject the null hypothesis of equal correlation matrix spectrum, we say that the HCM is empirically time invariant. It is not enough to have the same eigenvalues distribution since two matrices with no structural group features can still have the same spectrum. Indeed, the hidden correlation matrix should represent common features of the stocks. Figure \ref{fig:real_data_HCM_full} shows the Pearson correlation matrix, that we consider as a valuable comparison due to its widely usage in the financial community, against the HCM constructed with the SDT decomposition without removing the market factor. observations need to be made here. First of all, the Pearson correlation matrix has almost no structure. In fact, correlations are very small even for close to diagonal elements (same industry stocks). On the contrary, the clustering performance of the SDT decompositions improves considerably with respect to the Pearson correlation, even if, apart for some industries, a clear cut between sectorial groups is not evident. In particular, we note that stocks representing the Basic materials sector are clustered on the top left of the correlation matrix while the Financial industry in the bottom right of it. Other sectors appear to be more difficult to group in the middle of the matrix. 
		
\begin{figure}[H]			
	\centering
	\includegraphics[width=0.9\linewidth,height=0.25\textheight]{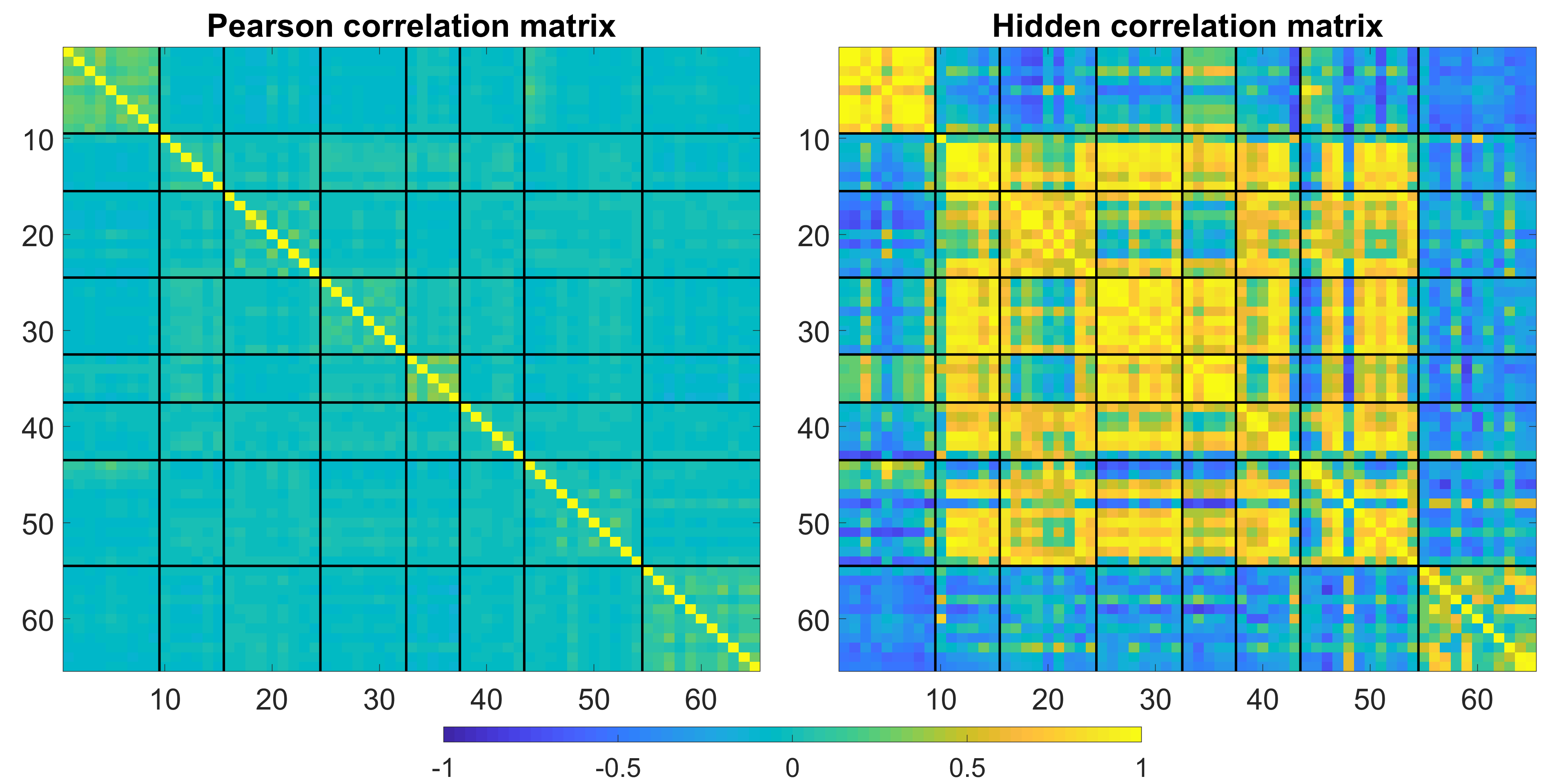}	
	\caption{Comparison between the Pearson correlation matrix and the HCM built from the SDT decomposition without the market mode. Black lines subdivide the correlation matrix in industry sectors, which are: Basic materials, Consumer goods, Services, Finance, Utilities, Technology, Industrial goods and Healthcare. Colours go from blue to yellow in the range [-1, 1], where -1 and 1 are the lowest and highest values of correlation empirically found.}
	\label{fig:real_data_HCM_nmkt}
\end{figure}

However, as we mentioned in the previous subsection, the first component is related to the market mode. Since the results can be severely affected by this common factor, we built the hidden correlation matrix removing it. Results are shown in figure \ref{fig:real_data_HCM_nmkt}. It is possible to notice that removing the market factor, the HCM shows more clearly the sectorial groups while the Pearson correlation matrix does not benefit from this procedure and the sectors are not visible any more. These results show that the hidden correlation matrix has huge structural features which are not hidden or netted out from different market dynamics. To strengthen this claim we finally run the same procedure over two non-overlapping samples and analyse the results via both visual inspection and statistical tests.

	\subsubsection{Time dependency analysis}
\begin{figure}[H]			
		\centering
		\includegraphics[width=0.9\linewidth,height=0.25\textheight]{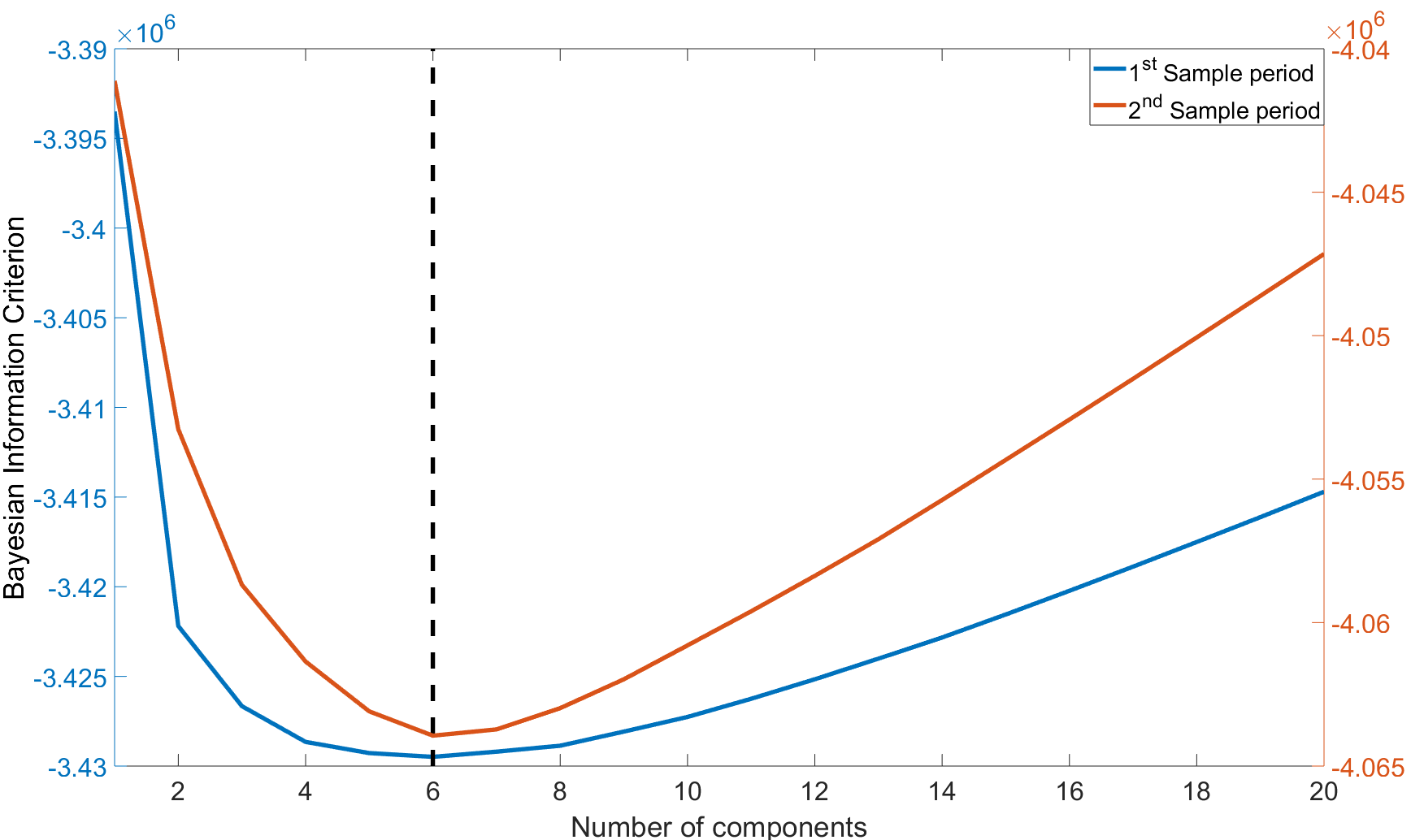}	
		\caption{Bayesian information criterion (BIC) for the number of correlation components of the SDT decomposition over the two non-overlapping samples. The plot shows a minimum for both sample periods at $6$.}
		\label{fig:real_data_BIC2}
	\end{figure}

	After the analysis of the structural features of the hidden correlation matrix, we study its behaviour over two non-overlapping sample periods. In particular, we split the tensor in two parts on the time dimension, generating two covariance tensors of dimension $65\times65 \times 75$. We fit the proposed SDT decomposition method to each tensor separately and build the two hidden correlation matrices. As it is possible to notice from figure \ref{fig:real_data_BIC2}, in both sample periods the number of correlation components for which the BIC is minimized is $6$. This is a first sign that the same number of relevant features is preserved over time.    

\begin{figure}[H]			
		\centering
	\includegraphics[width=0.9\linewidth,height=0.25\textheight]{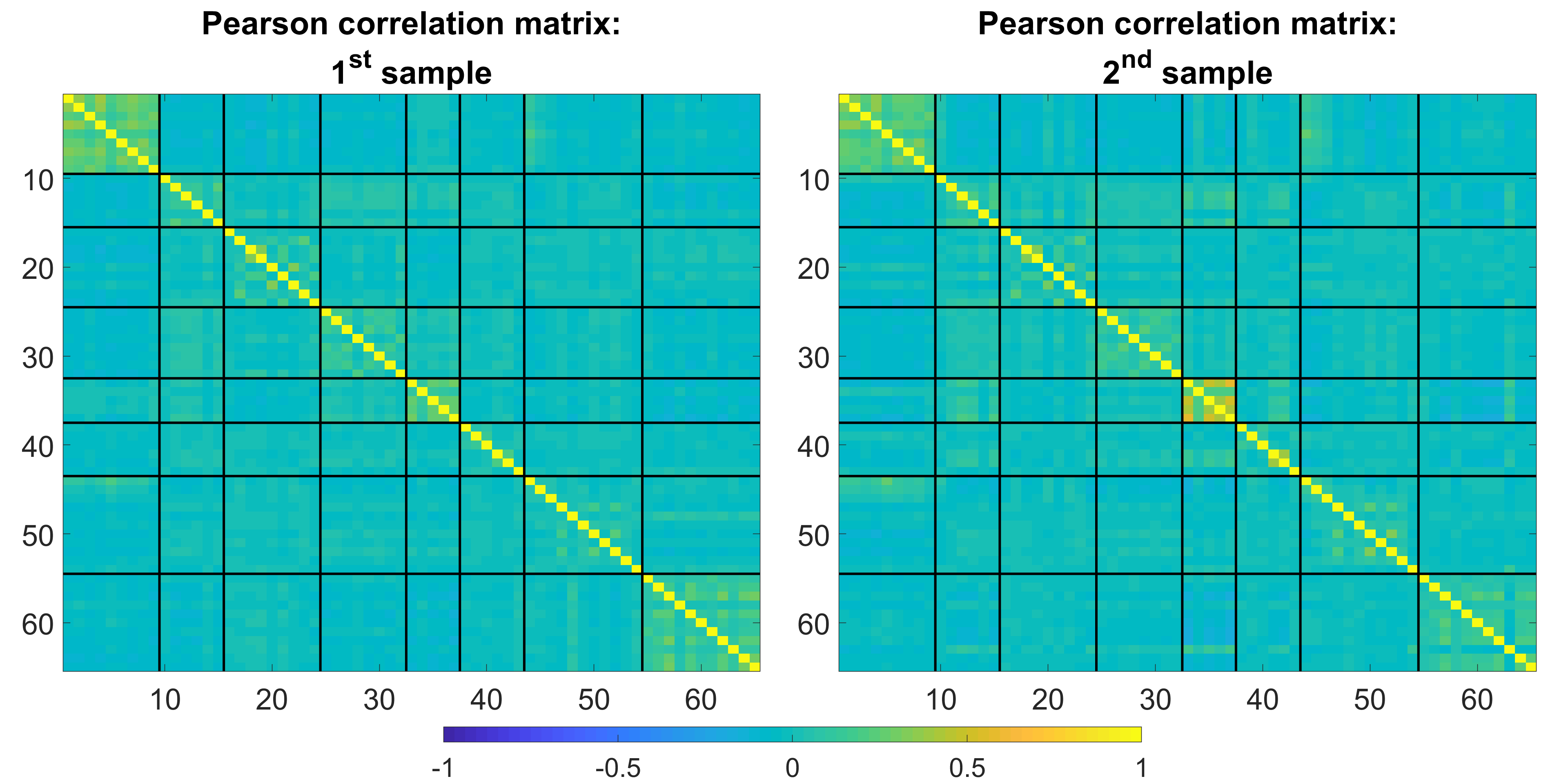}	
	\caption{Comparison between the Pearson correlation matrices computed over two non-overlapping samples without the market mode. Black lines subdivide the correlation matrix in industry sectors, which are: Basic materials, Consumer goods, Services, Finance, Utilities, Technology, Industrial goods and Healthcare. Colours go from blue to yellow in the range [-1, 1], where -1 and 1 are the lowest and highest values of correlation empirically found.}
	\label{fig:real_data_HCM_full2}
\end{figure}

\begin{figure}[H]			
		\centering
	\includegraphics[width=0.9\linewidth,height=0.25\textheight]{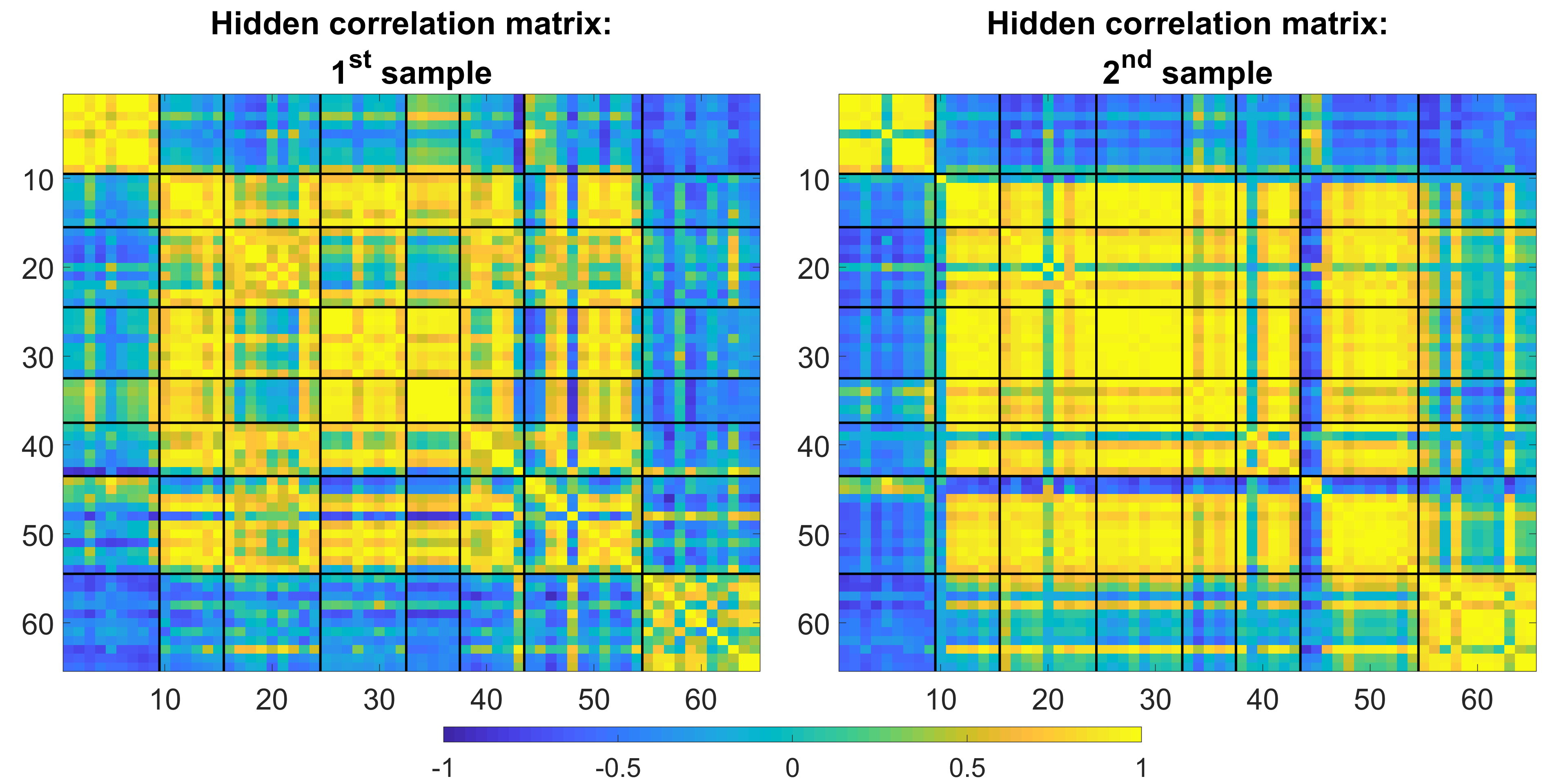}	
	\caption{Comparison between the HCMs built with the SDT components over two non-overlapping samples without the market mode. Black lines subdivide the correlation matrix in industry sectors, which are: Basic materials, Consumer goods, Services, Finance, Utilities, Technology, Industrial goods and Healthcare. Colours go from blue to yellow in the range [-1, 1], where -1 and 1 are the lowest and highest values of correlation empirically found.}
	\label{fig:real_data_HCM_nmkt2}
\end{figure}

As reported in the previous subsection, the market mode acts heavily on the correlation matrices. To eliminate this issue, we build both the Pearson and the hidden correlation matrix without the market mode to look at structural links which are not affected by this market. The results for the Person and Hidden correlation matrices are shown in figures \ref{fig:real_data_HCM_full2} and \ref{fig:real_data_HCM_nmkt2} respectively. The superiority of the hidden correlation matrix with respect to the Person correlation matrix is remarkable. In the latter, a structure emerges, while for the former it is even worsen and the correlations become negligible. 

The two sample HCMs have very similar structure and the same stocks clusters emerge. This a sign that a structural dependency matrix exists and that the proposed procedure is able to unveil it. 

To quantitatively asses what we showed through the visual inspection of the structural features of the two HCMs, we test if that structure is statistically similar over the two samples in terms of spectral similarity. Spectral similarity is the concept for which two matrices have the same eigenvalues. To statistically test for this feature, we implement two \textit{difference in distribution} tests for the spectrum of the two HCMs. If the distribution of the eigenvalues of the two matrices are not statistically different, i.e. a test is not able to reject the null of equal distribution, we can say that we have empirically similar correlation matrices \cite{Gera2018,HUANG2016}.\footnote{Recall that this is due to the fact that besides having statistically equivalent spectrum distributions, they show a non trivial block structure resembling market sectors.} In particular, we perform a Kruskal-Wallis test and a two-samples Kolmogorov-Smirnov non-parametric tests. The failure in rejecting the null hypothesis is in favour of the structure similarity of the matrices analysed. These tests, conceptually specify the following null and alternative hypothesis: 
	\[H_0:F(\lambda_{1,i})-F(\lambda_{2,i})=0 
	\]
	\[H_1:F(\lambda_{1,i})-F(\lambda_{2,i})\neq 0 
	\]

	where $F(\cdot)$ is the distribution of the data and $\lambda_{j,i}$ are the eigenvalues ($i=1,2,\dots,I$) of the $j$-th hidden correlation matrix. Specification of the two tests can be found in \cite{daniel1990applied}.
	Results are shown in Table \ref{test}. We can see that the null hypothesis cannot be rejected by both tests in both the cases in which the market mode is included or not, suggesting that the two sets of eigenvalues do come from the same distribution, even when a common factor, like the market mode, is removed.\footnote{We also run the tests using only the non-zero eigenvalues and the full rank specification, i.e. SDT decomposition with $65$ correlation components. The results are qualitatively equal.}
	
\begin{table}[H]
	\centering

	\begin{tabular}{|l|l|l|}
		\hline
		\textbf{Test}               & \textbf{{{With Market mode}} } & \textbf{{{Without Market mode}}} \\ \hline
		Kruskal-Wallis     &  0.9573  &  0.9351\\ \hline
		Kolmogorov-Smirnov & 1.0000  & 0.9995\\ \hline
	\end{tabular}
\caption{Test of similarity between the hidden correlation matrices eigenvalues. Reported values are the p-value of the test.}\label{test}
\end{table}
This analysis shows that the HCM has both the necessary features to be denoted as a structural time invariant correlation matrix, i.e. a non-trivial clustering structure and the statistical equivalence of eigenvalue distributions of the HCMs built over non-overlapping sample periods. This result allows us to consider the hidden correlation matrix as a good substitute of the correlation matrix with enhanced clustering and long-term characteristics. In relation to the other decomposition techniques, we only report that the Tucker decomposition has almost identical results with respect to the SDT model, while the PARAFAC showed poor results in both sample periods. 
	\section{Conclusion}\label{conclusion_sec}
In this paper, we proposed a way to construct a stable correlation matrix with emerging structural properties. In particular, we explored the factor components of the $3$-rd order covariance tensor of the S\&P100 with different decomposition techniques. We have proposed a new factorization technique, the SDT decomposition, which proved to have good performances with respect to the standard PARAFAC and Tucker decompositions. We showed via simulation that when a structural correlation matrix is present, we are able to unveil it from the tensor factorization. We then showed, using the that S\&P100 covariance tensor, that the dynamic component of the three factorizations analysed is a proxy of the VIX, confirming the reliability of these decompositions in disentangling dynamic component from static, cross-sectional components. The latter components are then used to build a correlation matrix that we named hidden correlation matrix, which is weakly dependent to the sample period used to estimate it. We appreciated the fact that the hidden correlation matrix constructed over two non-overlapping samples showed a non-trivial clustering structure and statistical similarity of the eigenvalue distributions. This makes this matrix a reliable proxy for the structural linkages between stocks. Such a matrix is readily applicable to portfolio construction when rebalancing is not possible and a structural dependence between stocks is necessary. Other possible applications are financial contagion, in which inferring the structural network from the data is fundamental, stress testing portfolios and forecasting cross-assets VaR for which the future dependencies should be as similar as possible to the current ones. In addition to these features, this approach results in being also a parsimonious tool to forecast the full correlation matrix in a future point in time since only the time factors have to be forecasted and then combined with the hidden correlation matrix.\\
In conclusion, besides the promising performance of the HCM presented in this paper, it is important to highlight again that the methodology developed is based on the assumption that a HCM exists. If the correlations structure changes permanently over time, the method is sub-optimal.\footnote{In case the dynamic change is given by a market shift, the HCM would still be stable and retrieving it via tensor decomposition feasible.} However, the HCM would still be more robust than a standard Pearson correlation matrix over the full sample period.       

\subsection*{Acknowledgements}
 We want to thank the ESRC Network Plus project 'Rebuilding macroeconomics'. We acknowledge support from Economic and Political Science Research Council (EPSRC) grant EP/P031730/1. We are grateful to NVIDIA corporation for supporting our research in this area with the donation of a GPU. We thank Bloomberg for providing the data.

      \bibliographystyle{unsrt}

      \bibliography{PaperCTP}	

\begin{thebibliography}{10}

\bibitem{Markowitz59}
Harry Markowitz.
\newblock {\em Portfolio Selection: Efficient Diversification of Investments}.
\newblock New York : Wiley, 1959.

\bibitem{aste2010correlation}
Tomaso Aste, W~Shaw, and Tiziana Di~Matteo.
\newblock Correlation structure and dynamics in volatile markets.
\newblock {\em New Journal of Physics}, 12(8):085009, 2010.

\bibitem{pozzi2012exponential}
Francesco Pozzi, Tiziana Di~Matteo, and Tomaso Aste.
\newblock Exponential smoothing weighted correlations.
\newblock {\em The European Physical Journal B}, 85(6):175, 2012.

\bibitem{musmeci2017multiplex}
Nicol{\`o} Musmeci, Vincenzo Nicosia, Tomaso Aste, Tiziana Di~Matteo, and Vito
  Latora.
\newblock The multiplex dependency structure of financial markets.
\newblock {\em Complexity}, 2017, 2017.

\bibitem{mantegna1999}
Rosario~N. Mantegna.
\newblock Hierarchical structure in financial markets.
\newblock {\em The European Physical Journal B-Condensed Matter and Complex
  Systems}, 11(1):193--197, 1999.

\bibitem{bartolozzi1}
Marco Bartolozzi, Christopher Mellen, Tiziana {Di Matteo}, and Tomaso Aste.
\newblock Multi-scale correlations in different futures markets.
\newblock {\em The European Physical Journal B-Condensed Matter and Complex
  Systems}, 58(2):207--220, 2007.

\bibitem{tumminello2010}
Michele Tumminello, Fabrizio Lillo, and Rosario~N. Mantegna.
\newblock Correlation, hierarchies, and networks in financial markets.
\newblock {\em Journal of Economic Behavior \& Organization}, 75(1):40--58,
  2010.

\bibitem{Schonbrodt2013}
Felix~D. Schönbrodt and Marco Perugini.
\newblock At what sample size do correlations stabilize?
\newblock {\em Journal of Research in Personality}, 47(5):609 -- 612, 2013.

\bibitem{gordon1998}
Gordon Y.~N. Tang.
\newblock The intertemporal stability of the covariance and correlation
  matrices of hong kong stock returns.
\newblock {\em Applied Financial Economics}, 8(4):359--365, 1998.

\bibitem{kao2018}
Chihwa Kao, Lorenzo Trapani, and Giovanni Urga.
\newblock Testing for instability in covariance structures.
\newblock {\em Bernoulli}, 24(1):740--771, 02 2018.

\bibitem{engel1999}
James Engel and Marianne Gizycki.
\newblock {Value at Risk: On the Stability and Forecasting of the
  Variance-covariance Matrix}.
\newblock RBA Research Discussion Papers rdp1999-04, Reserve Bank of Australia,
  May 1999.

\bibitem{lee1998}
Stephen Lee.
\newblock The inter-temporal stability of real estate returns: An empirical
  investigation.
\newblock Eres, European Real Estate Society (ERES), 1998.

\bibitem{piet1996}
Piet M.~A. Eichholtz.
\newblock The stability of the covariances of international property share
  returns.
\newblock {\em The Journal of Real Estate Research}, 11(2):149--158, 1996.

\bibitem{cheung1991}
Yan-Leung Cheung and Yan-Ki Ho.
\newblock The intertemporal stability of the relationships between the asian
  emerging equity markets and the developed equity markets.
\newblock {\em Journal of Business Finance \& Accounting}, 18(2):235--253,
  1991.

\bibitem{bro1998multi}
Rasmus Bro.
\newblock {\em Multi-way analysis in the food industry: models, algorithms, and
  applications}.
\newblock PhD thesis, K{\o}benhavns UniversitetK{\o}benhavns Universitet,
  LUKKET: 2012 Det Biovidenskabelige Fakultet for F{\o}devarer,
  Veterin{\ae}rmedicin og NaturressourcerFaculty of Life Sciences, LUKKET: 2012
  Institut for F{\o}devarevidenskabDepartment of Food Science, LUKKET: 2012
  Kvalitet og TeknologiQuality \& Technology, 1998.

\bibitem{henrion1994n}
Ren{\'e} Henrion.
\newblock N-way principal component analysis theory, algorithms and
  applications.
\newblock {\em Chemometrics and intelligent laboratory systems}, 25(1):1--23,
  1994.

\bibitem{kroonenberg2008applied}
Pieter~M Kroonenberg.
\newblock {\em Applied multiway data analysis}, volume 702.
\newblock John Wiley \& Sons, 2008.

\bibitem{grasedyck2013literature}
Lars Grasedyck, Daniel Kressner, and Christine Tobler.
\newblock A literature survey of low-rank tensor approximation techniques.
\newblock {\em GAMM-Mitteilungen}, 36(1):53--78, 2013.

\bibitem{kolda2009tensor}
Tamara~G Kolda and Brett~W Bader.
\newblock Tensor decompositions and applications.
\newblock {\em SIAM review}, 51(3):455--500, 2009.

\bibitem{anandkumar2014tensor}
Animashree Anandkumar, Rong Ge, Daniel Hsu, Sham~M Kakade, and Matus Telgarsky.
\newblock Tensor decompositions for learning latent variable models.
\newblock {\em Journal of Machine Learning Research}, 15:2773--2832, 2014.

\bibitem{faber2003recent}
Nicolaas Klaas~M Faber, Rasmus Bro, and Philip~K Hopke.
\newblock Recent developments in candecomp/parafac algorithms: a critical
  review.
\newblock {\em Chemometrics and Intelligent Laboratory Systems},
  65(1):119--137, 2003.

\bibitem{harshman1970foundations}
Richard~A Harshman.
\newblock Foundations of the parafac procedure: models and conditions for an"
  explanatory" multimodal factor analysis.
\newblock 1970.

\bibitem{carroll1970analysis}
J~Douglas Carroll and Jih-Jie Chang.
\newblock Analysis of individual differences in multidimensional scaling via an
  n-way generalization of “eckart-young” decomposition.
\newblock {\em Psychometrika}, 35(3):283--319, 1970.

\bibitem{tucker1963implications}
Ledyard~R Tucker.
\newblock Implications of factor analysis of three-way matrices for measurement
  of change.
\newblock {\em Problems in measuring change}, 122137, 1963.

\bibitem{tucker1964extension}
Ledyard~R Tucker.
\newblock The extension of factor analysis to three-dimensional matrices.
\newblock {\em Contributions to mathematical psychology}, 110119, 1964.

\bibitem{tucker1966some}
Ledyard~R Tucker.
\newblock Some mathematical notes on three-mode factor analysis.
\newblock {\em Psychometrika}, 31(3):279--311, 1966.

\bibitem{schwarz1978}
Gideon Schwarz.
\newblock Estimating the dimension of a model.
\newblock {\em Ann. Statist.}, 6(2):461--464, 03 1978.

\bibitem{akaike1974}
H.~{Akaike}.
\newblock A new look at the statistical model identification.
\newblock {\em IEEE Transactions on Automatic Control}, 19(6):716--723,
  December 1974.

\bibitem{CAVANAUGH1997}
Joseph~E. Cavanaugh.
\newblock Unifying the derivations for the akaike and corrected akaike
  information criteria.
\newblock {\em Statistics \& Probability Letters}, 33(2):201 -- 208, 1997.

\bibitem{bader2012matlab}
Brett~W Bader, Tamara~G Kolda, et~al.
\newblock Matlab tensor toolbox version 2.5.
\newblock {\em Available online, January}, 7, 2012.

\bibitem{Jolliffe}
Ian {Jolliffe}.
\newblock {\em Principal component analysis}.
\newblock Springer Verlag, New York, 2002.

\bibitem{matrix_factorization}
Carl~D. Meyer, editor.
\newblock {\em Matrix Analysis and Applied Linear Algebra}.
\newblock Society for Industrial and Applied Mathematics, Philadelphia, PA,
  USA, 2000.

\bibitem{appelof}
C.~J. Appellof and E.~R. Davidson.
\newblock Strategies for analyzing data from video fluorometric monitoring of
  liquid chromatographic effluents.
\newblock {\em Analytical Chemistry}, 53(13):2053--2056, 1981.

\bibitem{bro1997parafac}
Rasmus Bro.
\newblock Parafac. tutorial and applications.
\newblock {\em Chemometrics and intelligent laboratory systems},
  38(2):149--171, 1997.

\bibitem{anderson1999general}
CA~ANDERSON and R~HENRION.
\newblock A general algorithm for obtaining simple structure of core arrays in
  n-way pca with application to fluorometric data.
\newblock {\em Computational Statistics and Data Analysis}, 31:255--278, 1999.

\bibitem{andersen2003practical}
Charlotte~M{\o}ller Andersen and R~Bro.
\newblock Practical aspects of parafac modeling of fluorescence
  excitation-emission data.
\newblock {\em Journal of Chemometrics}, 17(4):200--215, 2003.

\bibitem{bro2006}
Rasmus Bro.
\newblock Review on multiway analysis in chemistry—2000–2005.
\newblock {\em Critical Reviews in Analytical Chemistry}, 36(3-4):279--293,
  2006.

\bibitem{acar2005modeling}
Evrim Acar, Seyit~Ahmet Camtepe, Mukkai~S Krishnamoorthy, and B{\"u}lent Yener.
\newblock Modeling and multiway analysis of chatroom tensors.
\newblock {\em ISI}, 2005:256--268, 2005.

\bibitem{acar2006}
Evrim Acar, Seyit~A. {\c{C}}amtepe, and B{\"u}lent Yener.
\newblock Collective sampling and analysis of high order tensors for chatroom
  communications.
\newblock In Sharad Mehrotra, Daniel~D. Zeng, Hsinchun Chen, Bhavani
  Thuraisingham, and Fei-Yue Wang, editors, {\em Intelligence and Security
  Informatics}, pages 213--224, Berlin, Heidelberg, 2006. Springer Berlin
  Heidelberg.

\bibitem{bader2007}
Brett W.~Bader, Richard Harshman, and Tamara G.~Kolda.
\newblock Temporal analysis of semantic graphs using asalsan.
\newblock pages 33--42, 10 2007.

\bibitem{bader2008}
Brett W.~Bader, Michael W.~Berry, and Murray Browne.
\newblock {\em Discussion tracking in enron email using PARAFAC}, pages
  147--163.
\newblock 01 2008.

\bibitem{kapteyn1986approach}
Arie Kapteyn, Heinz Neudecker, and Tom Wansbeek.
\newblock An approach ton-mode components analysis.
\newblock {\em Psychometrika}, 51(2):269--275, 1986.

\bibitem{de2000multilinear}
Lieven De~Lathauwer, Bart De~Moor, and Joos Vandewalle.
\newblock A multilinear singular value decomposition.
\newblock {\em SIAM journal on Matrix Analysis and Applications},
  21(4):1253--1278, 2000.

\bibitem{de2004dimensionality}
Lieven De~Lathauwer and Joos Vandewalle.
\newblock Dimensionality reduction in higher-order signal processing and
  rank-(r1, r2,…, rn) reduction in multilinear algebra.
\newblock {\em Linear Algebra and its Applications}, 391:31--55, 2004.

\bibitem{kiers2001three}
Henk~AL Kiers and Iven~Van Mechelen.
\newblock Three-way component analysis: Principles and illustrative
  application.
\newblock {\em Psychological methods}, 6(1):84, 2001.

\bibitem{kroonenberg1983three}
Pieter~M Kroonenberg.
\newblock {\em Three-mode principal component analysis: Theory and
  applications}, volume~2.
\newblock DSWO press, 1983.

\bibitem{cichocki2009nonnegative}
Andrzej Cichocki, Rafal Zdunek, Anh~Huy Phan, and Shun-ichi Amari.
\newblock {\em Nonnegative matrix and tensor factorizations: applications to
  exploratory multi-way data analysis and blind source separation}.
\newblock John Wiley \& Sons, 2009.

\bibitem{nickel2011three}
Maximilian Nickel, Volker Tresp, and Hans-Peter Kriegel.
\newblock A three-way model for collective learning on multi-relational data.
\newblock In {\em Proceedings of the 28th international conference on machine
  learning (ICML-11)}, pages 809--816, 2011.

\bibitem{kroonenberg1980principal}
Pieter~M Kroonenberg and Jan De~Leeuw.
\newblock Principal component analysis of three-mode data by means of
  alternating least squares algorithms.
\newblock {\em Psychometrika}, 45(1):69--97, 1980.

\bibitem{diffit}
Marieke~E. Timmerman and Henk A.~L. Kiers.
\newblock Three-mode principal components analysis: Choosing the numbers of
  components and sensitivity to local optima.
\newblock {\em British Journal of Mathematical and Statistical Psychology},
  53(1):1--16, 2000.

\bibitem{concordia}
Rasmus Bro and Henk A.~L. Kiers.
\newblock A new efficient method for determining the number of components in
  parafac models.
\newblock {\em Journal of Chemometrics}, 17(5):274--286, 2003.

\bibitem{Sena2006}
Marcelo~M. Sena, Marcello~G. Trevisan, and Ronei~J. Poppi.
\newblock Combining standard addition method and second-order advantage for
  direct determination of salicylate in undiluted human plasma by
  spectrofluorimetry.
\newblock {\em Talanta}, 68(5):1707 -- 1712, 2006.

\bibitem{higham2002computing}
Nicholas~J Higham.
\newblock Computing the nearest correlation matrix—a problem from finance.
\newblock {\em IMA journal of Numerical Analysis}, 22(3):329--343, 2002.

\bibitem{LKJ2009}
Daniel Lewandowski, Dorota Kurowicka, and Harry Joe.
\newblock Generating random correlation matrices based on vines and extended
  onion method.
\newblock {\em Journal of Multivariate Analysis}, 100(9):1989 -- 2001, 2009.

\bibitem{daniel1990applied}
W.W. Daniel.
\newblock {\em Applied Nonparametric Statistics}.
\newblock Duxbury advanced series in statistics and decision sciences.
  PWS-KENT, 1990.

\bibitem{Gera2018}
Ralucca Gera, L.~Alonso, Brian Crawford, Jeffrey House, J.~A. Mendez-Bermudez,
  Thomas Knuth, and Ryan Miller.
\newblock Identifying network structure similarity using spectral graph theory.
\newblock {\em Applied Network Science}, 3(1):2, Jan 2018.

\bibitem{HUANG2016}
Xu~Huang, Mansi Ghodsi, and Hossein Hassani.
\newblock A novel similarity measure based on eigenvalue distribution.
\newblock {\em Transactions of A. Razmadze Mathematical Institute}, 170(3):352
  -- 362, 2016.

\bibitem{kolda2006multilinear}
Tamara~Gibson Kolda.
\newblock Multilinear operators for higher-order decompositions.
\newblock Technical report, Sandia National Laboratories, 2006.

\bibitem{kolda2005higher}
Tamara~G Kolda, Brett~W Bader, and Joseph~P Kenny.
\newblock Higher-order web link analysis using multilinear algebra.
\newblock In {\em Data Mining, Fifth IEEE International Conference on}, pages
  8--pp. IEEE, 2005.

\bibitem{liu2008hadamard}
Shuangzhe Liu and G{\~o}tz Trenkler.
\newblock Hadamard, khatri-rao, kronecker and other matrix products.
\newblock {\em Int. J. Inf. Syst. Sci}, 4(1):160--177, 2008.

\end{thebibliography}
      
      \begin{appendices}

      \section{Tensor components and operations}\label{optens}
      
      As in matrix algebra, tensors can be evaluated in terms of their subparts, which in the matrix case are column and row vectors, while for tensor are called \textit{subarrays}. Subarrays are defined as the result of fixing a subset of indexes of the tensor. Fixing all but two indexes, we get what are called \textit{slices}. Slices are two-dimensional objects that are sections of the tensor. For a third order tensor $\T{X}$ we have that $X_{i::}$ are $i=1,2,...,I$ horizontal slices, $X_{:j:}$ are $j=1,2,...,J$ lateral slices while $X_{::k}$ are $k=1,2,...,K$ frontal slices. A graphical representation of slices is provided in figure \ref{fig:t3} below:
      
      \begin{center}
      	\begin{figure}[H]		
      		\centering	
      		\includegraphics[width=0.8\textwidth,height=0.2\textheight]{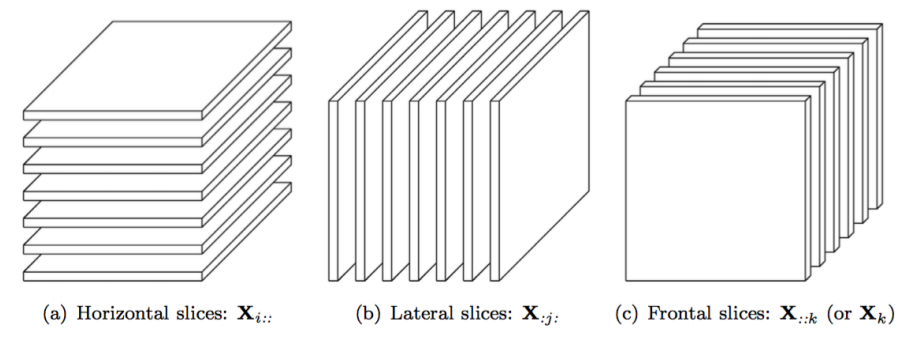}
      		\caption{Example of a $3rd$-order tensor slices}	
      		\label{fig:t3}
      	\end{figure}
      \end{center}
      
\begin{center}
      	\begin{figure}[H]			
      		\centering
      		\includegraphics[width=0.8\textwidth,height=0.2\textheight]{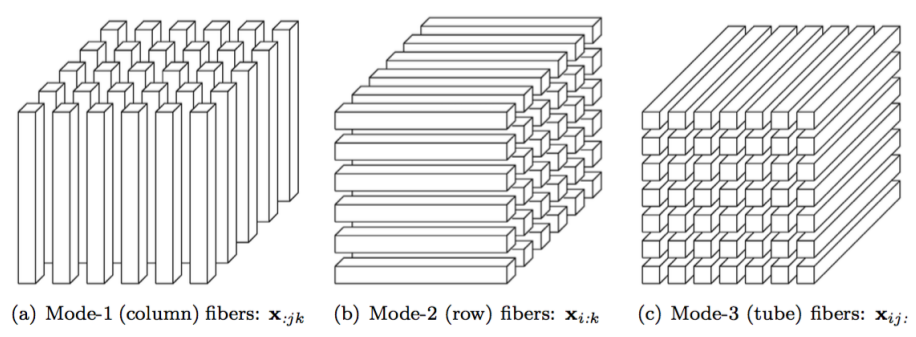}
      		\caption{Example of a $3rd$-order tensor modes fibers}	
      		\label{fig:t2}
      	\end{figure}
      \end{center}

      On the contrary, fixing all but one index we get \textit{fibers}. These objects are the higher order version of columns and rows of a matrix. For a third order tensor $\T{X}$ we have that $x_{:jk}$ is a column fiber, $x_{i:k}$ is a row fiber while $x_{ij:}$ is a tube fiber. figure \ref{fig:t2} gives a graphical explanation of these objects.   

     \section{Estimation algorithm}\label{als_algo}
     
     Let's write a general decomposition as  $F(\T{X})=\widehat{\T{X}},$ where $F(\cdot)$ is one of the three decomposition under examination. The minimization problem can be written as:
     
     \begin{equation}\label{opt}
     \min_{f_1,...,f_n}\parallel \T{X}-F(\T{X})\parallel_F,
     \end{equation}
     where $f_1,...,f_n$ are factors of the decomposition at hand. A sketch of the algorithm can be found below:
     \begin{algorithm}[H]
     	\caption{Alternating least squares}\label{ALS}
     	\begin{algorithmic}[1]
     		\State {Initialize by random generated factors $f_1,f_2,...,f_N$.}
     		\Repeat 
     		\For  {$\textit{i=1 to N}}$
     		\State {${\hat{f}_i=\argmin_{f_i}\parallel \T{X}-F(\T{X})_{\{f_1,...,f_{i-1},f_{i+1},...,f_N\}} \parallel_F}$}		
     		\EndFor
     		\State{$\varepsilon=\frac{\parallel \T{X}-F(\T{X})_{\{\hat{f}_1,\hat{f}_2,...,\hat{f}_N\}} \parallel_F}{\parallel \T{X}\parallel_F}$}
     		\Until {$\varepsilon<\alpha$ or Maximum iterations reached.}
     		\State {Return $\hat{f}_1,\hat{f}_2,...,\hat{f}_N$}
     		
     	\end{algorithmic}
     \end{algorithm}

\section{Operations on tensors}\label{optens2}
Multilinear algebra operations have been deeply analysed in the last decade. Among all the papers focusing on the topic, \cite{kolda2006multilinear,kolda2005higher,kolda2009tensor,de2000multilinear,anandkumar2014tensor,liu2008hadamard} are standard reference. Those papers show various operations involving tensors and matrices both formally and with examples. There are numerous operations which can be performed with tensors, but we present only the operations inherent to this work, i.e. matricization and $n$-mode product.
\subsubsection{Matricization}
The matricization of a tensor (also called unfolding) is the process of reshaping a tensor into a matrix. Take a tensor $\T{X}$ of size $I_1 \times I_{2}\cdots \times I_N $. Let the ordered sets $\mathcal{R} = {\{r_1,... , r_L\}}$ and $\mathcal{C} = \{c_1,... , c_M\}$ be a partitioning of $\mathcal{N}=\{1,2,\dots,N\}$. A matricized tensor is defined as:

\[X_{\{\mathcal{R} \times \mathcal{C}:I_\mathcal{N}\}} \in R^{J \times K}\] 
where
\[J=\prod_{n \in \mathcal{R} }I_{n}, \hspace{3pt} K=\prod_{n \in \mathcal{C} }I_{n}.\]
The set $\mathcal{R}$ maps the specified modes of the tensor onto the rows of the resulting matrix, while $\mathcal{C}$ on its columns.
The special case in which the set $\mathcal{R}$ is a singleton equal to $n$ and $\mathcal{C} = \{1,...,n-1,n+1,..., N\}$ is called \textbf{\emph{$n$-mode matricization}} and it is defined as: 
\[X_{\{\mathcal{R} \times \mathcal{C}:I_\mathcal{N}}\}\equiv X_{(n)}.\]
In this case, the fibers of $n$-th mode are aligned as the columns of the resulting matrix. 

\subsubsection{$n$-mode product}

The \emph{$n$-mode product} of the tensor $\T{X}$ of size $I_1 \times I_{2}\cdots \times I_N $ with the matrix $\M{V}$ of size $I_n \times J$ is denoted as
\begin{equation}\label{n_mode_prod}
\T{Y}=\T{X} \times_n \M{V}
\end{equation}

where the resulting tensor $\T{Y}$ is of size $I_1 \times \cdots \times I_{n-1} \times J	\times I_{n+1} \times \cdots \times I_N $. This can be expressed in terms of matricized tensors as:
\begin{equation}
\T{Y} = \T{X} \times_n \M{V} 
\quad \Leftrightarrow \quad 
\Mz{Y}{n} = \M{X}\Mz{Y}{n}.
\end{equation}
As in matrix algebra, the dimensions in which the operation is performed must commensurate and in the case more than one does, the dimension over which the operation need is performed must be specified. 

\subsection{Examples}\label{app_op}
Provided the definition of a tensor and of its sub-components, we provide some examples of these objects. Take as an example a tensor $\T{Y}$ of size $3 \times 4 \times 2$:

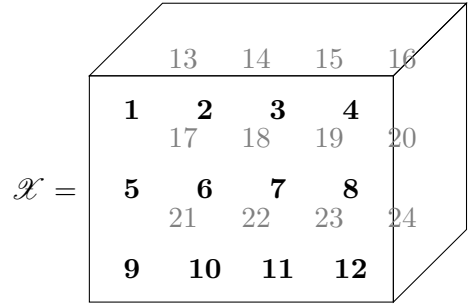
\begin{figure}[!h]
	\centering
	\begin{tikzpicture}[scale=.75,namenode/.style={scale=1}]

	\def\ix{4} %
	\def\iy{3} %
	\def\iz{2.5} %

	\def\corescale{0.75}
	\def\rx{\ix/\corescale}
	\def\ry{\iy/\corescale}
	\def\rz{\iz/\corescale}
	
	\coordinate (GFrontLowerLeft) at (0,0);
	\draw (GFrontLowerLeft) rectangle ++ (\rx,\ry);
	\begin{scope}[shift={(GFrontLowerLeft)},canvas is zx plane at y=\ry,rotate=90]
	\draw (0,0) rectangle ++ (\rx,\rz);
	\end{scope}
	\begin{scope}[shift={(GFrontLowerLeft)},canvas is zy plane at x=\rx,rotate=90]
	\draw (0,0) rectangle ++ (\ry,\rz);
	\end{scope}
	\node[namenode] at ($(GFrontLowerLeft)+(-0.15*\rx,.5*\ry)$)  {${\T{X}=}$};
	\node[namenode] at ($(GFrontLowerLeft)+(0.14*\rx,.85*\ry)$)  {$\textbf{1}$};
	\node[namenode] at ($(GFrontLowerLeft)+(0.38*\rx,.85*\ry)$)  {$\textbf{2}$};
	\node[namenode] at ($(GFrontLowerLeft)+(0.62*\rx,.85*\ry)$)  {$\textbf{3}$};
	\node[namenode] at ($(GFrontLowerLeft)+(0.86*\rx,.85*\ry)$)  {$\textbf{4}$};

	\node[namenode] at ($(GFrontLowerLeft)+(0.14*\rx,.5*\ry)$)  {$\textbf{5}$};
	\node[namenode] at ($(GFrontLowerLeft)+(0.38*\rx,.5*\ry)$)  {$\textbf{6}$};
	\node[namenode] at ($(GFrontLowerLeft)+(0.62*\rx,.5*\ry)$)  {$\textbf{7}$};
	\node[namenode] at ($(GFrontLowerLeft)+(0.86*\rx,.5*\ry)$)  {$\textbf{8}$};

	\node[namenode] at ($(GFrontLowerLeft)+(0.14*\rx,.15*\ry)$)  {$\textbf{9}$};
	\node[namenode] at ($(GFrontLowerLeft)+(0.38*\rx,.15*\ry)$)  {$\textbf{10}$};
	\node[namenode] at ($(GFrontLowerLeft)+(0.62*\rx,.15*\ry)$)  {$\textbf{11}$};
	\node[namenode] at ($(GFrontLowerLeft)+(0.86*\rx,.15*\ry)$)  {$\textbf{12}$};

	
	
	\node[namenode] at ($(GFrontLowerLeft)+(0.14*\rx,0.85*\ry, -.70*\rz)$)  {\textcolor{gray}{$13$}};
	\node[namenode] at ($(GFrontLowerLeft)+(0.38*\rx,0.85*\ry, -.70*\rz)$)  {\textcolor{gray}{$14$}};
	\node[namenode] at ($(GFrontLowerLeft)+(0.62*\rx,0.85*\ry, -.70*\rz)$)  {\textcolor{gray}{$15$}};
	\node[namenode] at ($(GFrontLowerLeft)+(0.86*\rx,0.85*\ry, -.70*\rz)$)  {\textcolor{gray}{$16$}};

	\node[namenode] at ($(GFrontLowerLeft)+(0.14*\rx,.5*\ry, -.70*\rz)$)  {\textcolor{gray}{$17$}};
	\node[namenode] at ($(GFrontLowerLeft)+(0.38*\rx,.5*\ry, -.70*\rz)$)  {\textcolor{gray}{$18$}};
	\node[namenode] at ($(GFrontLowerLeft)+(0.62*\rx,.5*\ry, -.70*\rz)$)  {\textcolor{gray}{$19$}};
	\node[namenode] at ($(GFrontLowerLeft)+(0.86*\rx,.5*\ry, -.70*\rz)$)  {\textcolor{gray}{$20$}};

	\node[namenode] at ($(GFrontLowerLeft)+(0.14*\rx,.15*\ry, -.70*\rz)$)  {\textcolor{gray}{$21$}};
	\node[namenode] at ($(GFrontLowerLeft)+(0.38*\rx,.15*\ry, -.70*\rz)$)  {\textcolor{gray}{$22$}};
	\node[namenode] at ($(GFrontLowerLeft)+(0.62*\rx,.15*\ry, -.70*\rz)$)  {\textcolor{gray}{$23$}};
	\node[namenode] at ($(GFrontLowerLeft)+(0.86*\rx,.15*\ry, -.70*\rz)$)  {\textcolor{gray}{$24$}};
	
	\end{tikzpicture}
	\caption{3rd-order tensor $\T{Y}$ of size $3 \times 4 \times 2$}
	\label{fig:tesnor_example}
\end{figure}	

This tensor can be represented through slices or fibers. Among the different representations of a tensor, frontal slices are the most used ones since they make the comparison with matrices easier. The two \textbf{frontal} slices (or modes) of the tensor $\T{Y}$ are:

\[X_{::1}=\left( \begin{array}{cccc}
1&4  &7  &10 \\
2&  5& 8&11\\
3& 6& 9&12\end{array} \right) \hspace{10pt}
X_{::2}=\left( \begin{array}{cccc}
13&16  &19  &22 \\
14& 17& 20&23\\
15& 18& 21&24\end{array} \right),\]

the four \textbf{lateral} slices of $\T{X}$ are:

\[X_{:1:}=\left( \begin{array}{cccc}
1&13 \\
5&  17\\
9& 21\end{array} \right) \hspace{3pt}
X_{:2:}=\left( \begin{array}{cccc}
2 & 14  \\
6 & 18 \\
10 & 22 \end{array} \right) \hspace{3pt}
X_{:3:}=\left( \begin{array}{cccc}
3&15\\
7& 19\\
11&23\end{array} \right) \hspace{3pt} X_{:4:}=\left( \begin{array}{cccc}
4  & 16\\
8 &  20\\
12 & 24\end{array} \right),\]

while the three \textbf{horizontal} slices of $\T{X}$ are:	
\[X_{1::}=\left( \begin{array}{cccc}
1&4  &7  &10 \\
13&  16& 16& 19\end{array} \right) \hspace{2pt}
X_{2::}=\left( \begin{array}{cccc}
2 & 5  & 8  & 11 \\
14 & 17 & 20 & 23 \end{array} \right)
\hspace{2pt}
X_{:1:}=\left( \begin{array}{cccc}
3&6  &9  &12 \\
15& 18& 21&24\end{array} \right).\]

Examples on fibers are numerous so we show only a few. A mode-1 fiber (\textbf{column} fiber) of the tensor $\T{X}$ can be given by:
\[X_{(:,1,1)}= \left( \begin{array}{c}
1 \\
2\\
3\end{array} \right) \hspace{10pt} or  \hspace{10pt} X_{:32}= \left( \begin{array}{c}
19 \\
20\\
21\end{array} \right), \]

examples of mode-2 fibers (\textbf{row} fiber) are:

\[X_{1:1}=\left( \begin{array}{cccc}
1&4  &7  &10 \\
\end{array} \right) \hspace{10pt} or  \hspace{10pt} X_{4:2}=\left(\begin{array}{cccc}
15& 18& 21&24\end{array}\right),\]
while a mode-3 fiber (\textbf{tube} fiber) can be represented as:
\[X_{11:}=\left( \begin{array}{cc}
1 &13 \\
\end{array} \right) \hspace{10pt} or  \hspace{10pt} X_{32:}=\left(\begin{array}{cc}
8& 20\end{array}\right).\]
Defining a tensor in terms of its sub-components make  the interpretation of the basic operations with tensors easier and in some circumstances makes some numerical computations more efficient. 

\subsection{Operations}\label{app-op}
\subsubsection*{Matricization}
Take the tensor $\T{X}$ of figure \ref{fig:tesnor_example}, a general matricization as specified in section \ref{optens}, can be given by $\mathcal{R}=\{3,1\}$ and $\mathcal{C}=\{2\}$, then:
\[X_{(\{3,1\} \times\{2\}:\{3,4,2\})}=\left( \begin{array}{cccc}
1&4 &7 &10 \\
13& 16& 19& 22 \\
2& 5& 8& 11 \\
14& 17& 20& 23 \\
3& 6& 9& 12 \\
15& 18& 21& 24\end{array} \right)\]

While the n-mode matricization (in the $1$-st mode) of the tensor $\T{X}$ is given by:
\[X_{(\{1\}\times \{2,3\}:3\times4\times2)}=X_{(1)}=\tilde{X}=\left( \begin{array}{cccccccc}
1&4  &7  &10&13&16&19&22 \\
2&  5& 8&11&14&17&20&23\\
3& 6& 9&12&15&18&21&24\end{array} \right).\]

\subsubsection*{n-mode product}

Take again the tensor $\T{X}$ of figure \ref{fig:tesnor_example} and $V=\left( \begin{array}{cccc}
1&2  &3 \\
4& 5& 6\\\end{array} \right).$ 
The product $\T{Y} =\T{X} \times_1\M{V}$ results in:
\[Y_{::1}=\left( \begin{array}{cccc}
14&32  &50&68 \\
32& 77& 122&167 \\\end{array} \right)\]
\[Y_{::2}=\left( \begin{array}{cccc}
86&104  &122&140 \\
212& 257& 302& 347\\\end{array} \right).\]

\section{Correlation}\label{Corr_sec}
In financial economics, the reference measure of association is the Pearson correlation, which is a linear form of association between random variables, generally applied to stock log-returns. Log-returns of stock $i$ at time $t$ are defined as follows:
\begin{equation}
r_{i,t}=ln(P_{i,t})-ln(P_{i,t-1}),
\end{equation}
where $P_{i,t}$ is the price of stock $i$ at time $t$. \\
Let $\E[r_i]$ denote the expected value of the stock $i$ log-returns. The variance of $r_i$ is defined as:
\begin{equation}Var(r_i)=\sigma_{r_i}^2=\E[(r_i-\E[r_i])^2]\end{equation}
while the covariance between a pair of stocks $i$ and $j$ in terms of log-returns $r_i$ and $r_j$ is defined as:
\begin{equation} \label{cov_eq1}
Cov(r_i,r_j)=\sigma_{r_i,r_j}=[(r_i-\E[r_i])(r_j-\E[r_j])].\end{equation}
Given this formulation we can define the Pearson correlation coefficient between two stocks as:
\begin{equation}corr(r_i,r_j)=\rho_{r_i,r_j}=\frac{Cov(r_i,r_j)}{\sqrt{Var(r_i)Var(r_j)}}.\end{equation}
It is possible to rewrite the previous formulae in matrix form. In particular, let's $\textbf{e}$ being a random vector containing the log-returns of several stocks, we can write \ref{cov_eq1} as:	
\begin{equation}
Var(\mathbf{e})= \Sigma= \E[(\mathbf{e}-\E[\mathbf{e}])(\mathbf{e}-\E[\mathbf{e}])'].
\end{equation}
Now, let's $\Omega$ be the correlation matrix associated to corr($\textbf{e}$) and $D$ a diagonal matrix such that $D_{kk}=\sqrt{Var(\mathbf{e_{k}})}$. We can then define the variance-covariance matrix $\Sigma$ as:
\begin{equation}\label{cov_eq}
\Sigma=D \Omega D,
\end{equation}
and
\begin{equation}
\Omega=D^{-1}\Sigma D^{-1}.
\end{equation}
Every entry of the $\Omega$ matrix is associated with the correlation between two random variables, i.e. $\Omega_{ij}$ is the correlation between $\mathbf{e_i}$ and $\mathbf{e_j}$. Despite its easiness of computation and interpretation, not all sample correlation matrices constructed via pairwise correlation are correlation matrices. In fact, to define a matrix as \textit{correlation matrix}, it is not enough to have a symmetric matrix with bounded values between $-1$ and $+1$ and with all $1s$ on the main diagonal. A correlation matrix must fulfil an additional property, which is to be positive semi-definite. All correlation matrices are positive semi-definite but not all sample correlation matrices are guaranteed to have this property. Take as an example the following matrix  $O$:
\[
O=\begin{bmatrix}
1	& -0,6 &	0,8\\
-0,6 &	1 &	0,8\\
0,8	& 0,8 &	1
\end{bmatrix}.
\]

A positive semi-definite matrix has non-negative eigenvalues. Nevertheless, the three eigenvalues (in increasing order)  of the matrix A are  -0.47, 1.60 and 1.87. 
This example provides evidence of a symmetric matrix, with values in the interval $[-1,+1]$ and values equal to $1$ on the diagonal which is not a correlation matrix. 

\section{Simulation results: Additional plots}\label{sim_res_plots}

\begin{figure}[H]		
	\centering	
	\includegraphics[width=0.8\linewidth,height=0.25\textheight]{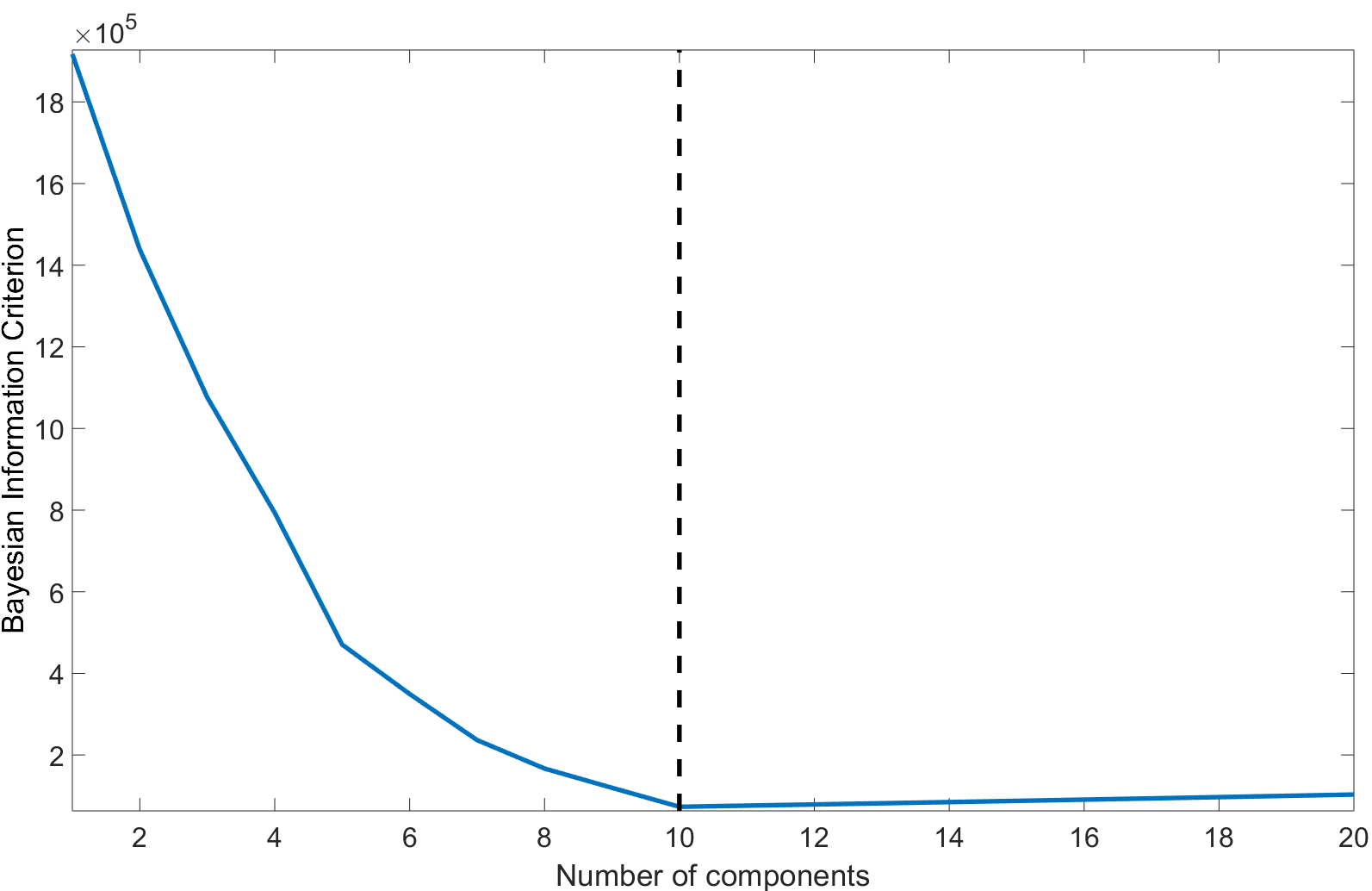}	
	\caption{The Bayesian information criterion (BIC) for the number of correlation components of the SDT decomposition. The plot shows a minimum at $10$ components.}	
	\label{fig:sim_BIC}	
\end{figure}

\begin{figure}[H]		
	\centering	
	\includegraphics[width=0.8\linewidth,height=0.25\textheight]{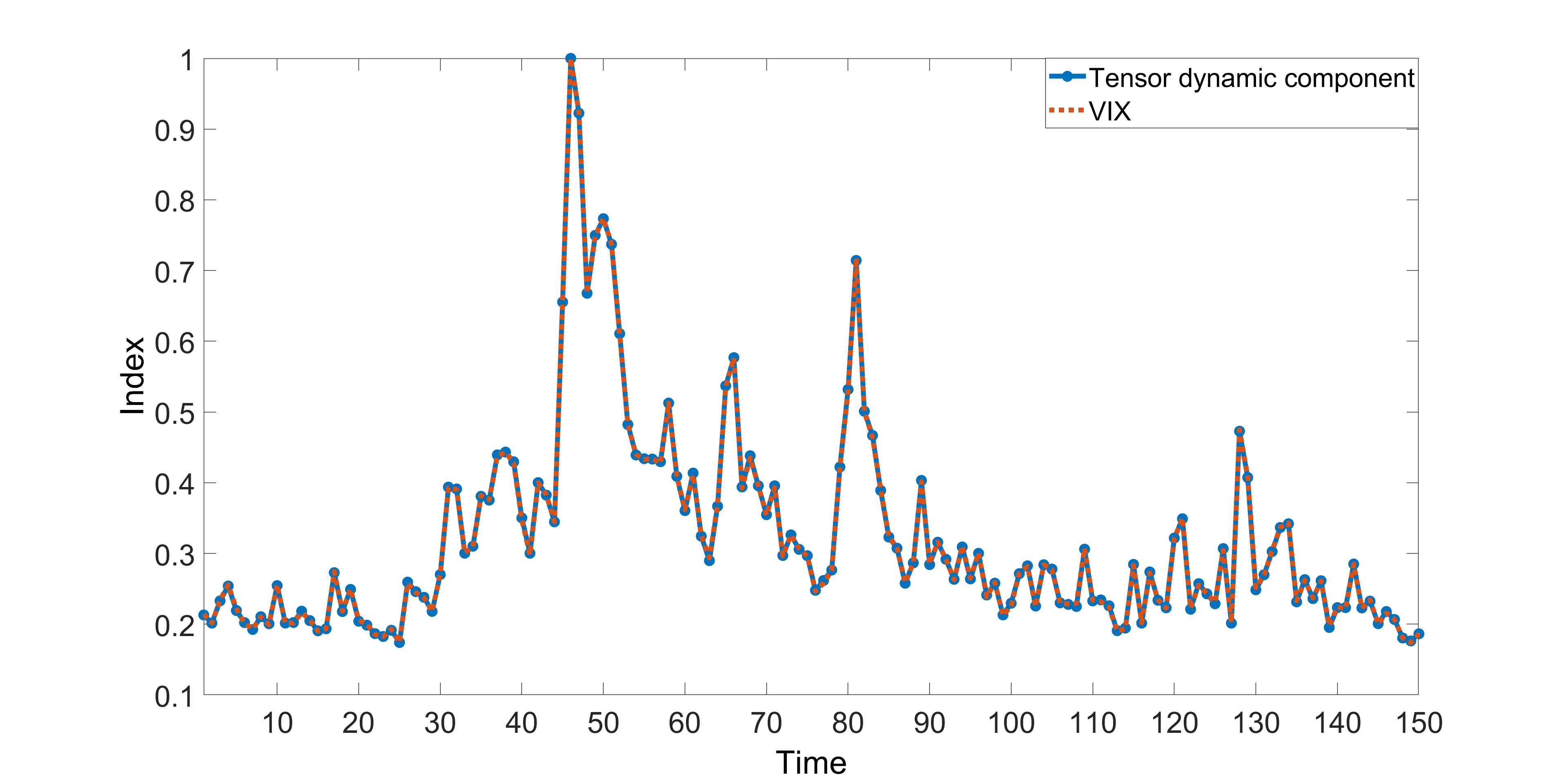}		
	\caption{Dynamic component retrieved by the SDT decomposition against and the VIX time series used to construct the tensor.}
	\label{fig:sim_time}	
\end{figure}
As it is possible to notice from figure \ref{fig:sim_time}, the decomposition is able to identify almost perfectly the time component.  
In fact, is almost impossible to spot any difference.

\begin{figure}[H]		
	\centering	
	\includegraphics[width=0.8\linewidth,height=0.25\textheight]{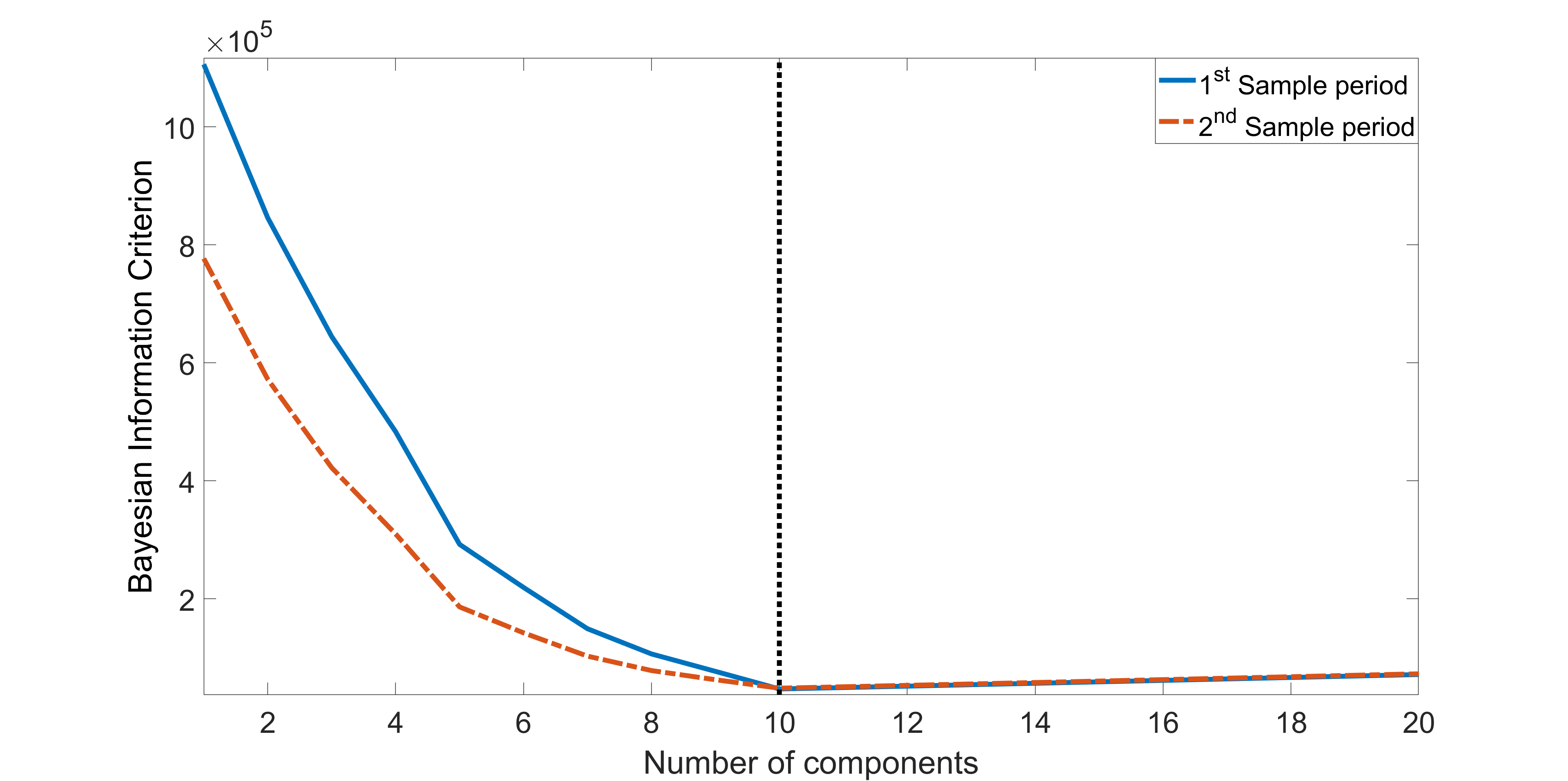}		
	\caption{The Bayesian information criterion (BIC) for the number of correlation components of the SDT decomposition in the two samples. The plot shows a minimum at $10$ components for both samples.}
	\label{fig:sim_BIC2}	
\end{figure}
      In figure \ref{fig:sim_BIC2} we can see that according to the BIC, the number of components is in both sample periods is 10, the number used to build the covariance tensor.

  \begin{figure}[H]
  		\centering	
  	\includegraphics[width=0.8\linewidth,height=0.25\textheight]{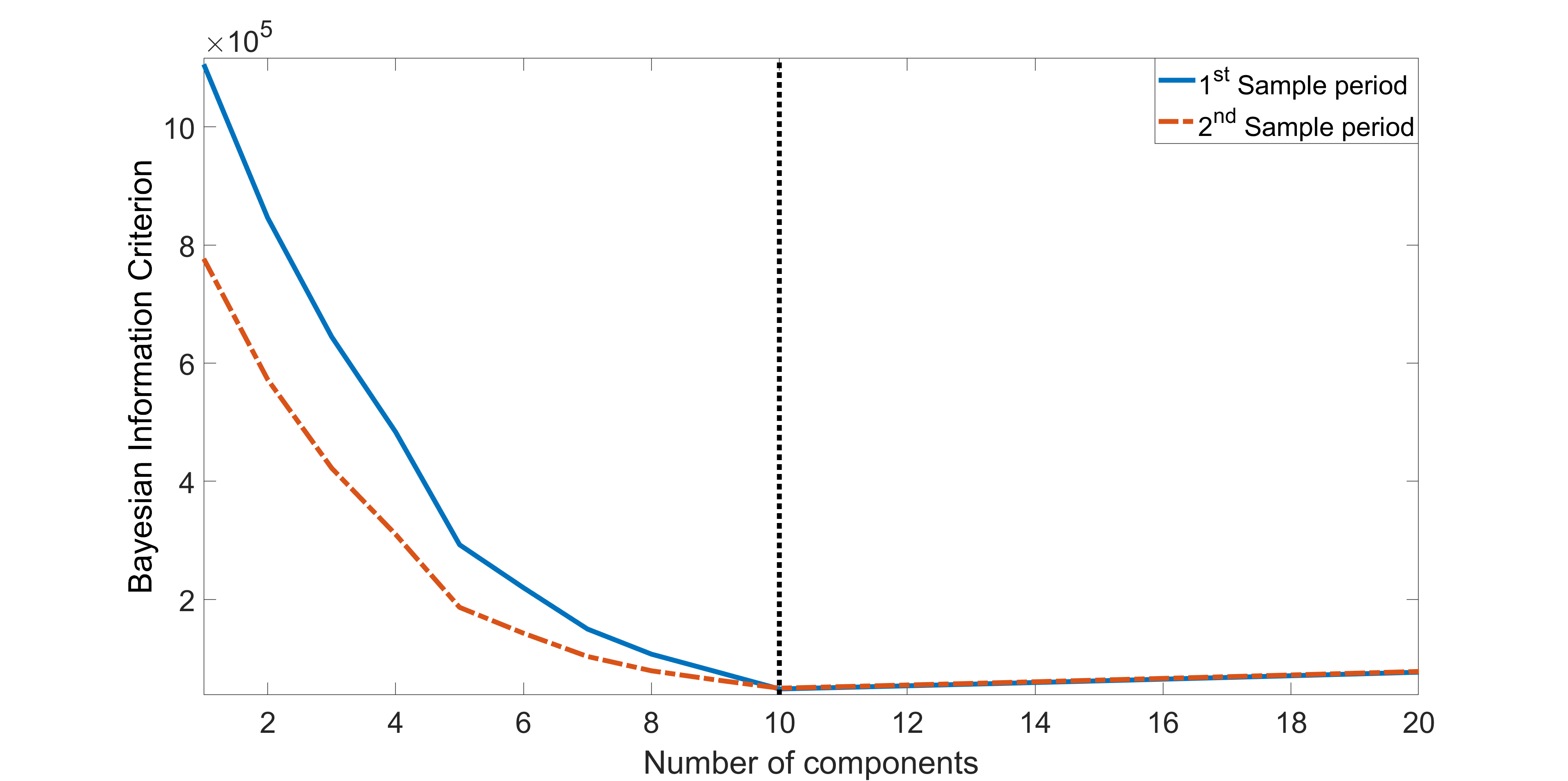}
  
  	\caption{The Bayesian information criterion (BIC) for the number of correlation components of the Tucker decomposition. The plot shows a minimum at $10$ components. }\label{fig:sim_HCM22}
  \end{figure}
  We can see from figure \ref{fig:sim_HCM22} that the results related to the number of correlation factors found by the BIC for the Tucker decomposition are almost identical to the SDT approach.

  \begin{figure}[H]

  		\centering	
  	\includegraphics[width=0.8\linewidth,height=0.25\textheight]{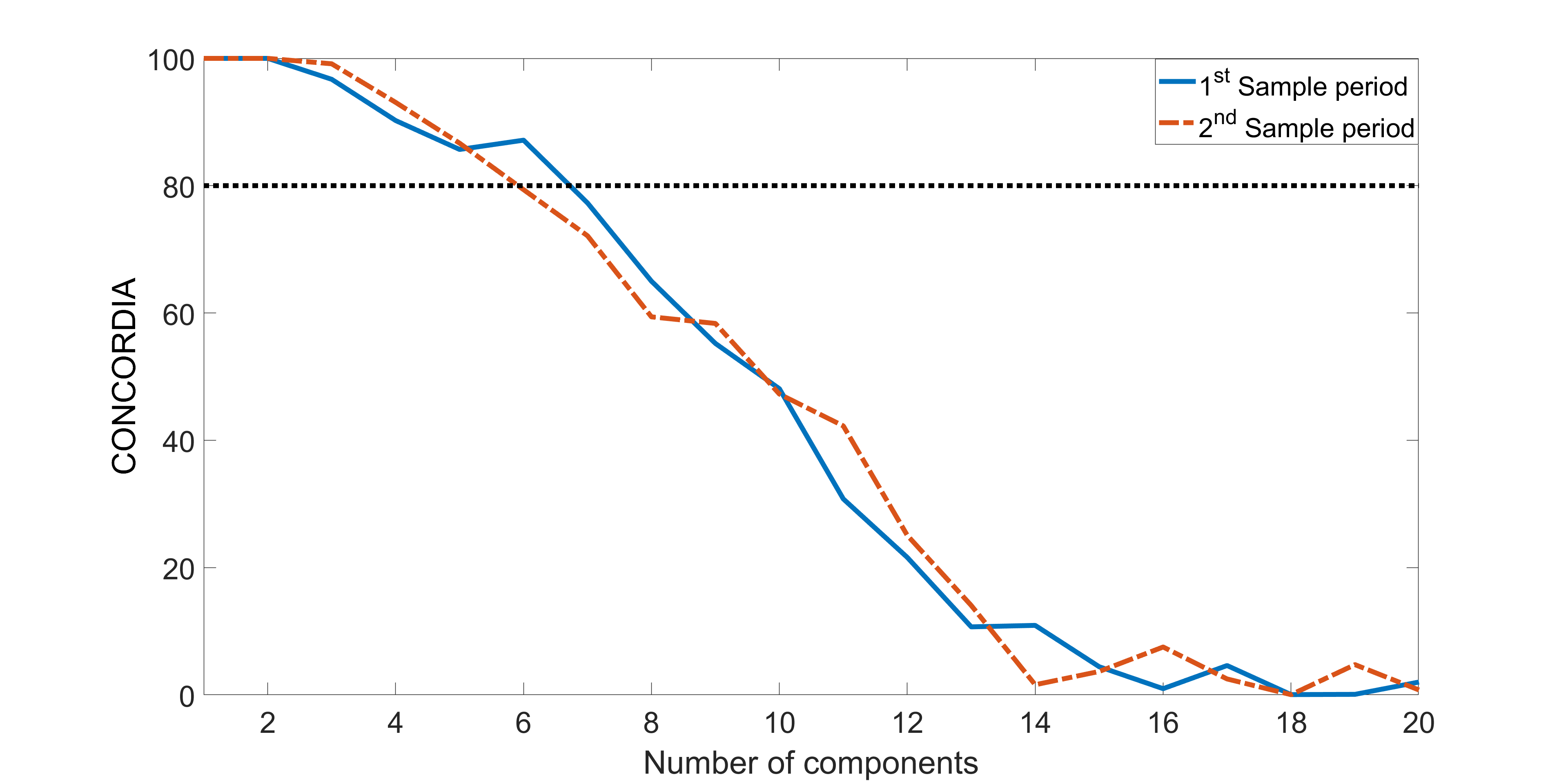}

  	\caption{The CONCORDIA criterion for the number of correlation components of the PARAFAC decomposition in the two samples. The plot shows optimal number of components at $6$ and $5$ in the two samples respectively.}\label{fig:sim_CONCORDIA}
  \end{figure}
      
      \section{Application results: Additional plots}\label{real_res_plots}

      	\begin{figure}[H]		
      		\centering	
      		\includegraphics[width=0.8\linewidth,height=0.25\textheight]{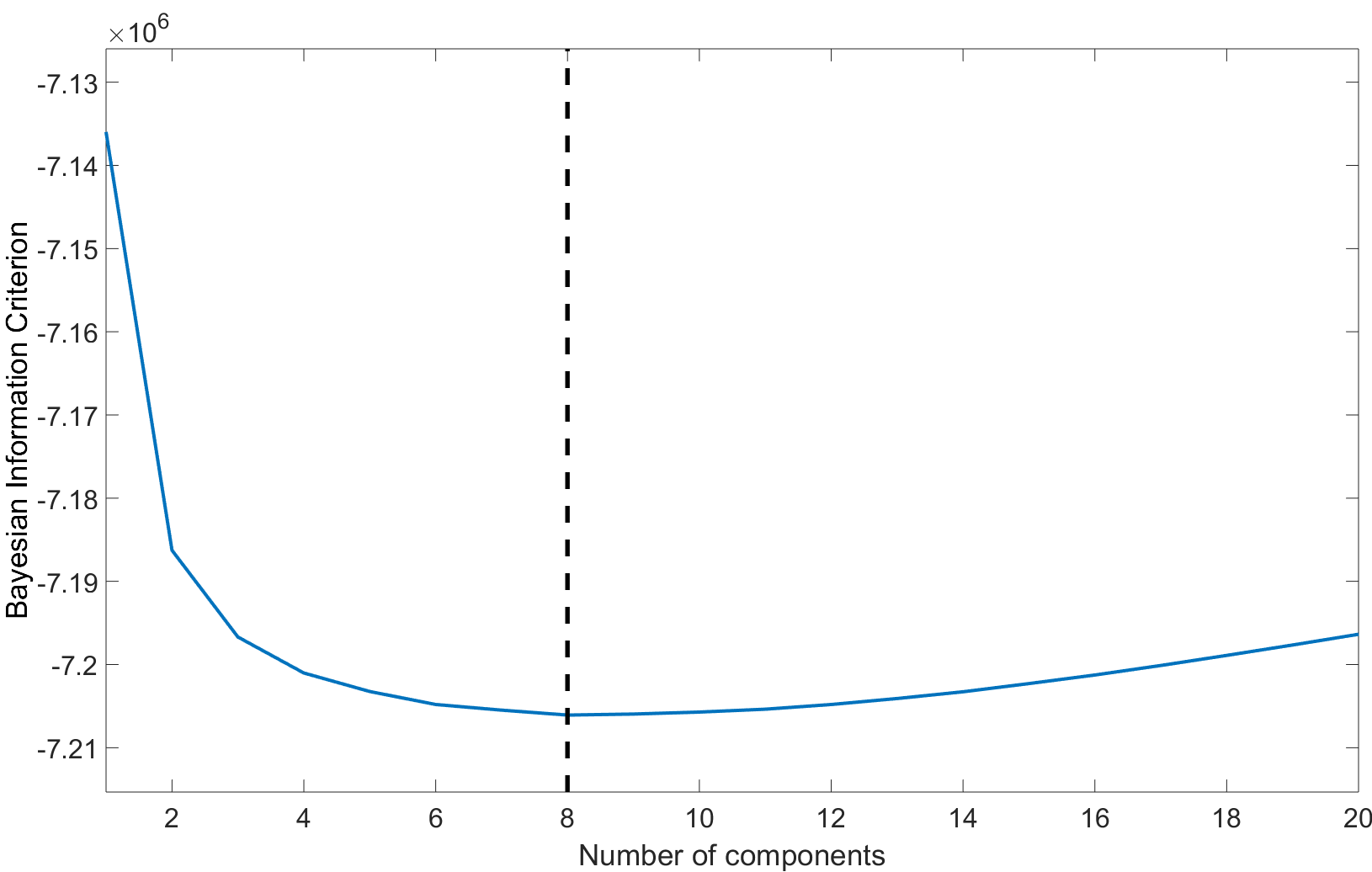}	
      		\caption{Bayesian information criterion (BIC) for the number of correlation components of the Tucker and SDT decompositions. The plot shows a minimum for both the model at $8$.}	
      		\label{fig:emprirical_IC}	
      	\end{figure}

      \begin{figure}[H]		
      	\centering	
      	\includegraphics[width=0.8\linewidth,height=0.25\textheight]{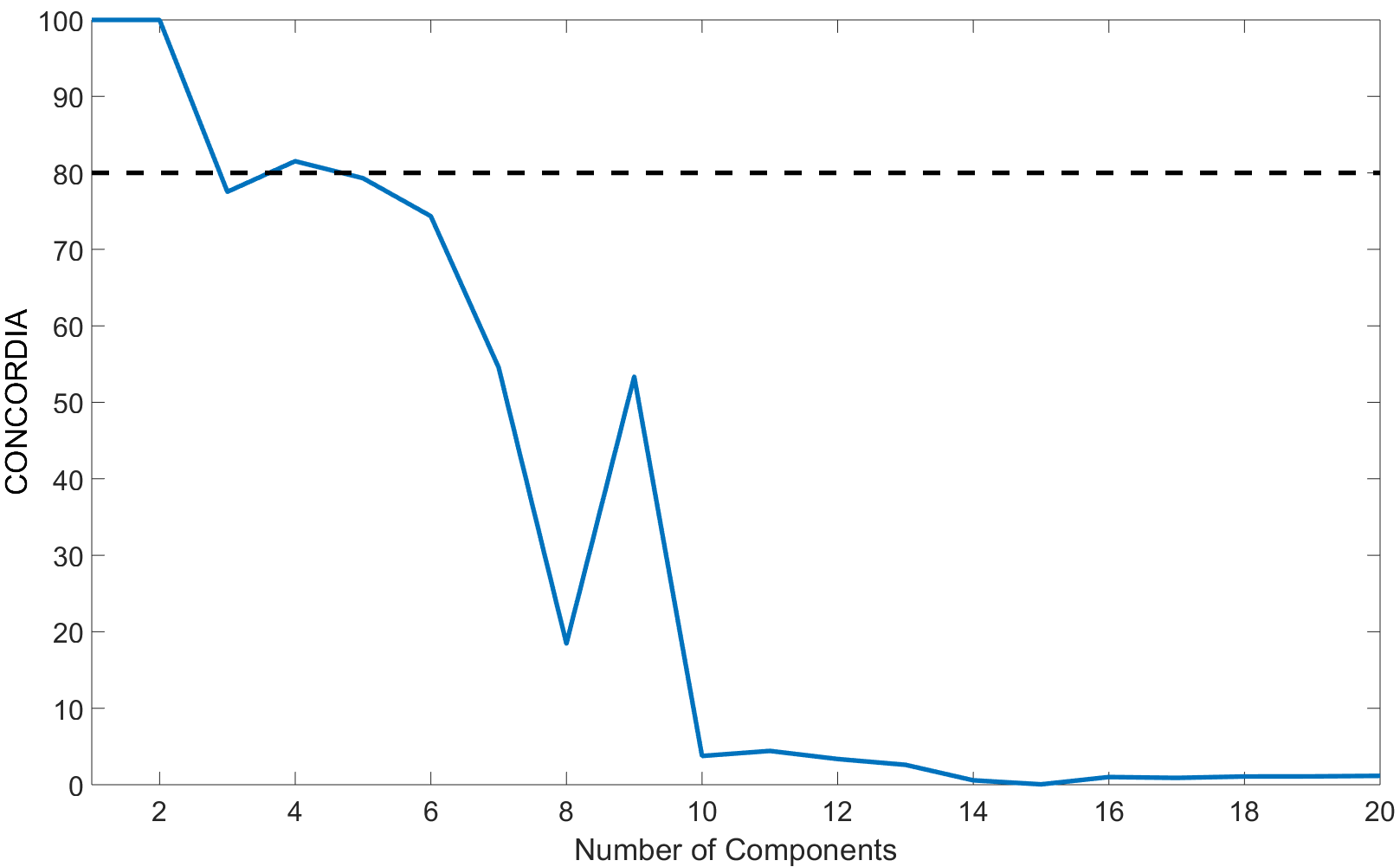}		
      	\caption{CONCORDIA criterion for the PARAFAC model. The plot shows two acceptable solutions with $2$ and $4$ components.}	
      	\label{fig:emprirical_CONCORDIA}
      \end{figure}
      
      As it is possible to notice from figure \ref{fig:emprirical_CONCORDIA}, the minimum threshold of $80\%$ is reached only for $2$ specifications (apart the trivial $1$ factor PARAFAC). As the general procedure requires we select the $4$ factors specification since it is the one with the higher number of components.

      \section{Static components}\label{stat_comps}
\begin{figure}[H]		
      	\centering	
      	\includegraphics[width=1\linewidth,height=0.4\textheight]{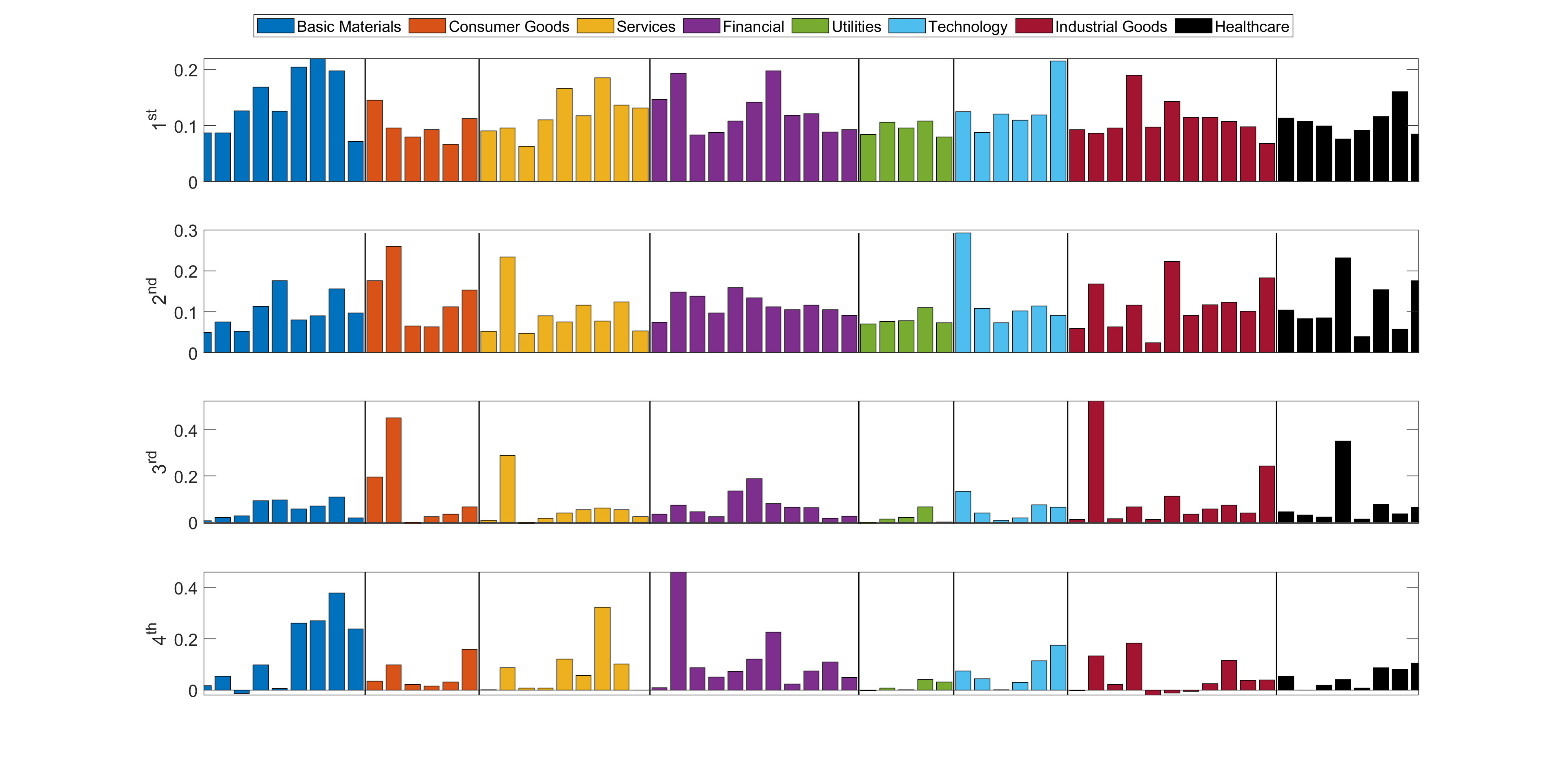}
      	\caption{The first 4 factor components of the PARAFAC decomposition. Different colours are associated to different industries, which are: Basic materials, Consumer goods, Services, Finance, Utilities, Technology, Industrial goods and Healthcare. Each sub-plot shows a factor component in increasing order from the top to the bottom.}
      	\label{fig:factorsP}		
      \end{figure}

      In this appendix we analyse the static components which are represented by the hidden space induced by the stock returns covariances. Figure \ref{fig:factorsP} depicts the $4$ correlation components of the PARAFAC decomposition. As it is possible to notice from the plot, the first factor loading exhibit an high and homogeneous value for all the stocks across different industries, while the other factor, especially $3^{rd}$ and $4^{th}$, depicts the heterogeneity across different market sectors.

\begin{figure}[H]	
      	\centering		
      	\includegraphics[width=1\linewidth,height=0.45\textheight]{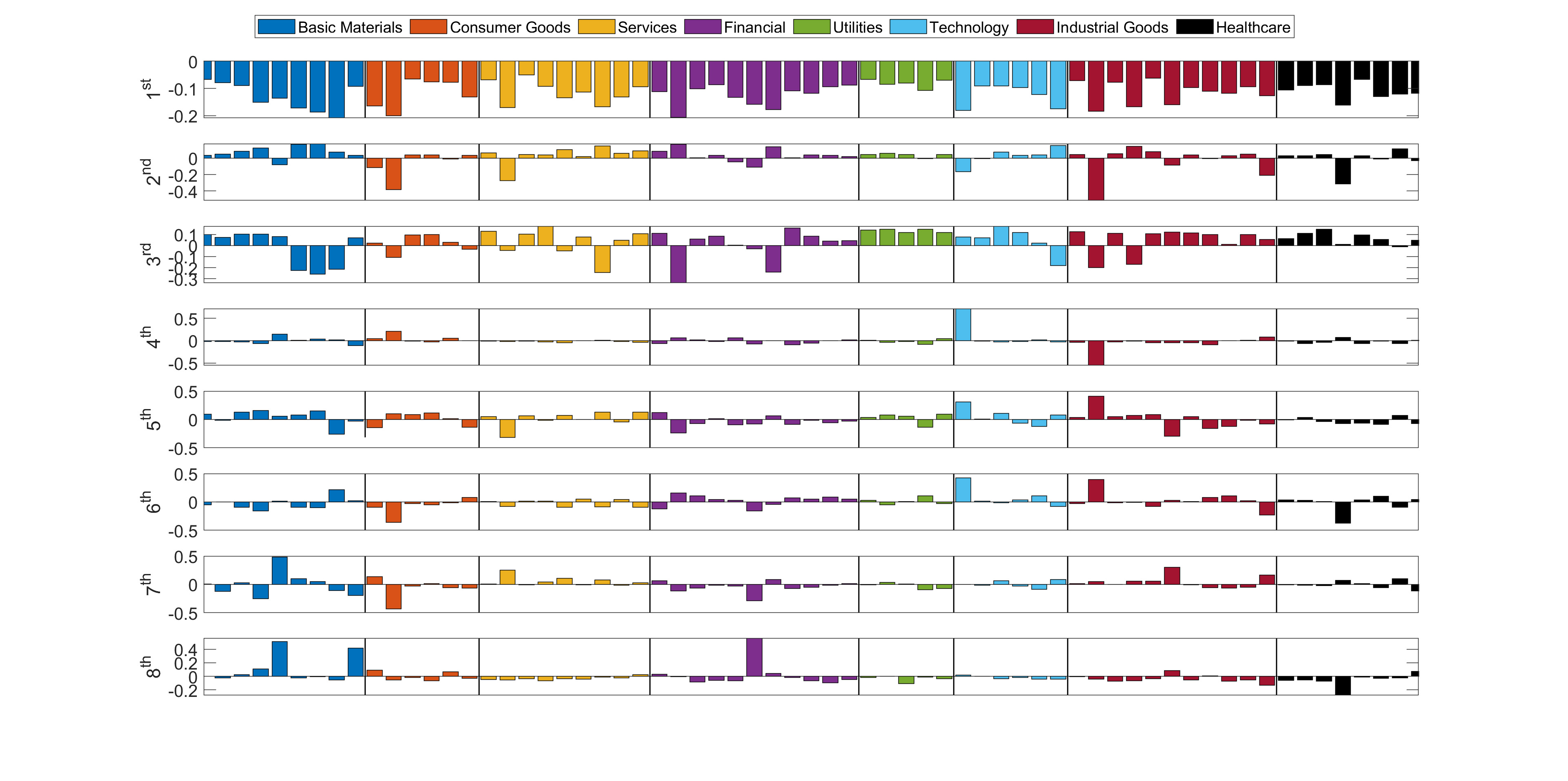}	
      	\caption{The first 8 factor components of the Tucker decomposition. Different colours are associated to different industries, which are: Basic materials, Consumer goods, Services, Finance, Utilities, Technology, Industrial goods and Healthcare. Each sub-plot shows a factor component in increasing order from the top to the bottom.}
      	\label{fig:factorsT}	
      \end{figure}

\begin{figure}[H]		
      	\centering	 	
      	\includegraphics[width=1\linewidth,height=0.45\textheight]{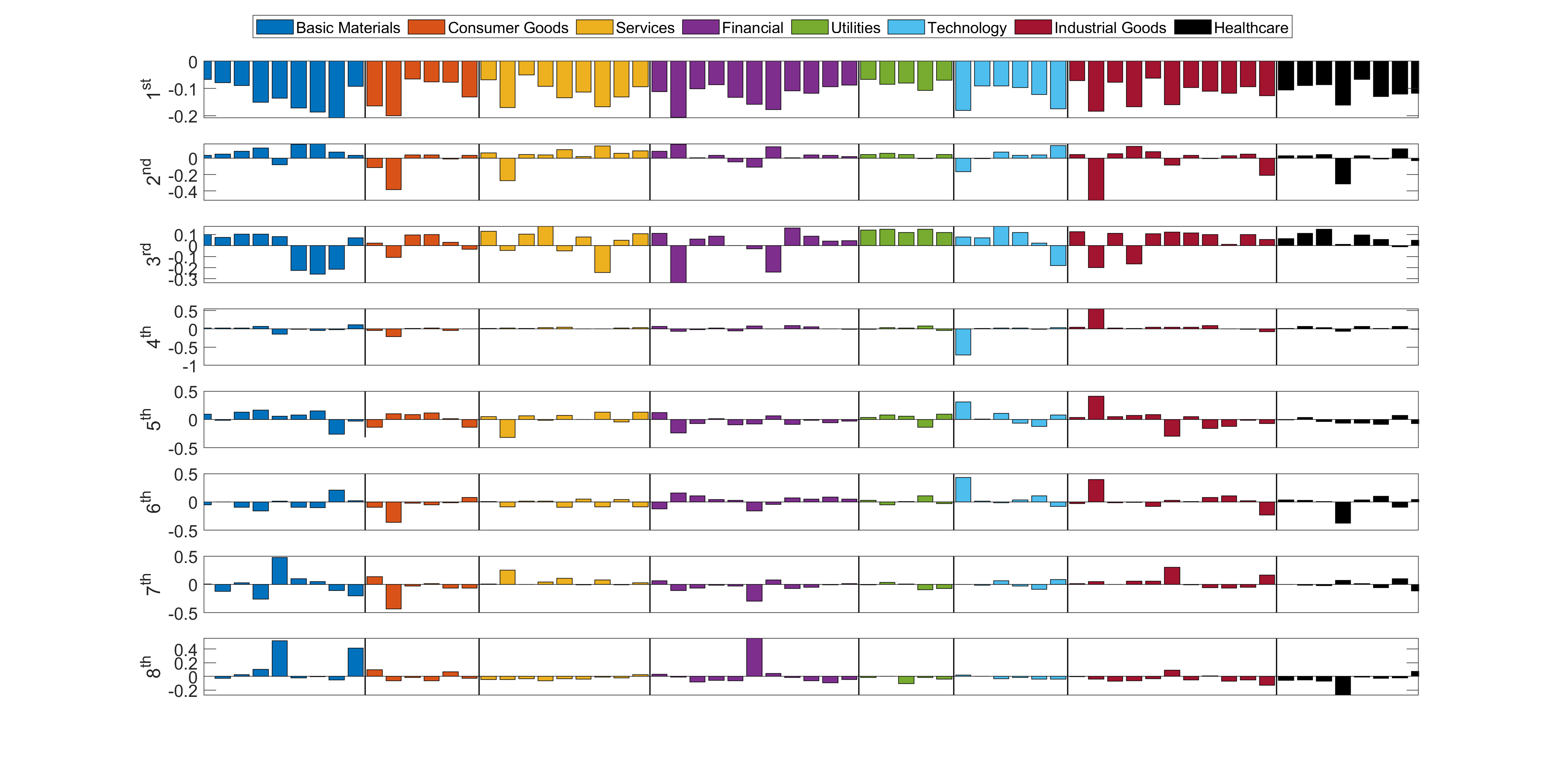}	
      	\caption{The first 8 factor components of the SDT decomposition. Different colours are associated to different industries, which are: Basic materials, Consumer goods, Services, Finance, Utilities, Technology, Industrial goods and Healthcare. Each sub-plot shows a factor component in increasing order from the top to the bottom.}	
      	\label{fig:factorsD}
      \end{figure}

      With respect to the Tucker and SDT decompositions, we can notice from figures \ref{fig:factorsT} and \ref{fig:factorsD} that the $8$ components are almost identical if not for imperceptible differences and the sign of the $4^{th}$ component. For this reason we will group the comments together. It is clear to see that the first component strongly affects all the industries in the same direction and in an homogeneous way. This means that this factor is common to all the sectors and it is usually attributed to the market component\footnote{The same argument is valid for the first component of the PARAFAC decomposition.}. The other components show some sort of fingerprint of the different industries even if a clear-cut to disentangle the features of each industry, apart some exception, is complicated to set. However, the important information we have to take from this analysis is that the first factor component is homogeneous among all assets, and represents the market mode.

      \end{appendices}     
\end{document}